\documentclass[showpacs,amsmath,amssymb,aps,preprint,nofootinbib]{revtex4-1}

\usepackage{graphicx}
\usepackage{dcolumn}
\usepackage{color} 
\usepackage{float}
\allowdisplaybreaks
\usepackage{hyperref}
\hypersetup{%
colorlinks=true,%
linkcolor=blue,
citecolor=blue,
urlcolor=magenta
}
\newcommand{\bpartial}{\stackrel{\leftrightarrow}{\partial}}
\newcommand{\rpartial}{\stackrel{\rightarrow}{\raisebox{-0.0ex}{$\partial$}}}
\newcommand{\lpartial}{\stackrel{\leftarrow}{\raisebox{-0.0ex}{$\partial$}}}
\def\gtsim{\mathrel{\hbox{\raise0.2ex
\hbox{$>$}\kern-0.75em\raise-0.9ex\hbox{$\sim$}}}}
\def\ltsim{\mathrel{\hbox{\raise0.2ex
\hbox{$<$}\kern-0.75em\raise-0.9ex\hbox{$\sim$}}}}

\begin{document}


\title{Exploring CP violation and vanishing electric dipole moment of the electron in a scale-invariant general 2HDM}

\author{Nguyen Dang Bao Nhi$^{1}$}
\email{dang.bao.nhi.nguyen@cern.ch}
\author{Eibun Senaha$^{2,3}$}
\email{eibunsenaha@vlu.edu.vn}
\affiliation{$^1$Institute of Particle and Nuclear Physics,
Faculty of Mathematics and Physics, Charles University in Prague, V Hole\v{s}ovi\v{c}k\'ach 2, 180 00 Praha 8, Czech Republic}
\affiliation{$^2$Subatomic Physics Research Group, Science and Technology Advanced Institute, Van Lang University, Ho Chi Minh City, Vietnam}
\affiliation{$^3$Faculty of Applied Technology, Van Lang School of Technology, Van Lang University, Ho Chi Minh City, Vietnam}

\date{\today}

\begin{abstract}
We investigate a scale-invariant general two-Higgs-doublet model with explicit CP violation. Using the Gildener-Weinberg method, we analyze the Higgs mass spectrum and couplings at the one-loop level, focusing on CP violation originating from both the Higgs potential and the Yukawa sector. We show that, due to the flatness condition, both CP-even and CP-odd mixings between the 125~GeV Higgs boson and the heavier Higgs states arise only radiatively. As a result, the Higgs couplings to gauge bosons and fermions remain Standard Model-like, in agreement with current LHC constraints.
We also find that the magnitude of CP violation from the Higgs potential depends on the new complex Yukawa couplings and heavy Higgs masses, due to the flatness condition. In the case where only an additional top Yukawa coupling (\(\rho_{tt}\)) is present, the electron electric dipole moment is directly proportional to \(\mathrm{Im}\rho_{tt}^2\). Furthermore, a nontrivial cancellation region can occur for a specific Higgs mass spectrum. We extend our analysis to scenarios with an additional complex electron Yukawa coupling, identifying the conditions under which the electron electric dipole moment is suppressed or vanishes.
\end{abstract}


\maketitle


\section{Introduction}\label{sec:intro}
Since the discovery of the Higgs boson, its properties have been gradually revealed, with current measurements from the LHC exhibiting excellent agreement with the predictions of the Standard Model (SM)~\cite{ATLAS:2022vkf,CMS:2022dwd}. Despite these experimental successes, the fundamental origin of electroweak symmetry breaking (EWSB) remains an open question. In the SM, EWSB is induced by a negative squared mass term in the tree-level Higgs potential. However, current data do not preclude alternative scenarios, such as radiative EWSB~\cite{Coleman:1973jx}, in which classical scale invariance is preserved at tree level and the electroweak scale arises dynamically via quantum corrections~\cite{Sher:1988mj}.

Although such a mechanism is, in principle, viable within the SM, standard perturbative analyses suggest that the electroweak vacuum becomes unstable due to the large top-quark contribution, rendering the potential unbounded from below.\footnote{See Ref.~\cite{Steele:2012av} for a resummation-based study that claims vacuum stability.}

The well-known shortcomings of the SM, including its inability to explain dark matter and the baryon asymmetry of the Universe, strongly motivate the existence of physics beyond the SM, though its energy scale remains unknown. Among the simplest extensions is the two-Higgs-doublet model (2HDM), which introduces a second scalar doublet~\cite{Branco:2011iw}. The 2HDM offers a well-motivated framework for electroweak baryogenesis (EWBG)~\cite{Kuzmin:1985mm} (see also Refs.~\cite{Rubakov:1996vz,*Funakubo:1996dw,*Riotto:1998bt,*Trodden:1998ym,*Bernreuther:2002uj,*Cline:2006ts,*Morrissey:2012db,*Konstandin:2013caa,*Senaha:2020mop}), accommodating both a strongly first-order electroweak phase transition and new sources of CP violation (CPV).

At present, the strength of CP violation is tightly constrained by null results from searches for the electron electric dipole moment (EDM), with the most stringent bounds reported by the ACME~\cite{ACME:2018yjb} and JILA~\cite{Roussy:2022cmp} experiments. These constraints have significantly restricted the viable parameter space for successful EWBG in 2HDMs with a softly broken $Z_2$ symmetry. On the other hand, the general 2HDM (g2HDM), in which no symmetry is imposed on the model, still allows for viable CPV sources. In particular, Ref.~\cite{Fuyuto:2017ewj} demonstrated that a complex extra top Yukawa coupling ($\rho_{tt}$) can generate the observed baryon asymmetry (see also Ref.~\cite{Aiko:2025tbk} for a recent study). Moreover, a cancellation mechanism involving a complex electron Yukawa coupling ($\rho_{ee}$) can suppress the electron EDM even in the presence of sizable CPV~\cite{Fuyuto:2019svr}.

In this work, we explore the \textit{scale-invariant general two-Higgs-doublet model} (SI-g2HDM)~\cite{Inoue:1979nn,Funakubo:1993jg,Takenaga:1993ux,Takenaga:1995ms,Sher:1996ib,Lee:2012jn,Hill:2014mqa,Fuyuto:2015jha,Lane:2018ycs,Lane:2019dbc,Eichten:2022vys}, focusing on CP violation arising from both the scalar and Yukawa sectors. Radiative EWSB is analyzed using the Gildener-Weinberg (GW) formalism~\cite{Gildener:1976ih}, which enables a consistent perturbative treatment of loop-induced effects on Higgs couplings and their phenomenological implications, including contributions to the electron EDM.

It is found that at tree level, the flatness conditions in the GW approach ensure the absence of CP mixing in the neutral Higgs mass matrix. At one-loop level, however, CP violation in the Higgs sector becomes correlated with that in the Yukawa sector through the GW flat direction. Since both CP-even and -odd mixings between the 125 GeV Higgs boson and the heavier neutral scalars arise only at the loop level, the Higgs couplings to gauge bosons and fermions remain consistent with current LHC measurements.

We also find that, when only the additional top Yukawa coupling $\rho_{tt}$ is nonzero, the dominant CP-violating contribution to the electron EDM is proportional solely to $\mathrm{Im}\, \rho_{tt}^2$. Our analysis reveals that the electron EDM can vanish 
in the parameter region where the CP-even-like Higgs boson is heavier than the CP-odd-like one, and the charged Higgs mass is nearly degenerate with the CP-odd Higgs mass (i.e., the custodially symmetric case). 
In contrast, such a cancellation region does not arise in the twisted custodially symmetric scenario, 
where the non-SM-like CP-even Higgs boson is nearly degenerate with the charged Higgs boson. 
If, in addition, a nonzero electron Yukawa coupling is introduced via the parametrization 
$\rho_{ee} = -r \rho_{tt}^* (y_e / y_t)$~\cite{Fuyuto:2019svr}, 
where \( r \) is a real parameter and $y_t~(y_e)$ denotes the SM top (electron) Yukawa coupling, 
the electron EDM remains proportional to $\mathrm{Im}\, \rho_{tt}^2$. 
In this case, a nontrivial cancellation region appears in both the custodially and twisted custodially symmetric 
mass spectra for $r = \mathcal{O}(1)$.

The remainder of this paper is organized as follows. In Sec.~\ref{sec:model}, we present the SI-g2HDM and study its scalar sector at both tree and one-loop levels, with emphasis on CP-violating effects. Numerical results are also discussed. In Sec.~\ref{sec:EDMs}, we analyze the electron EDM, focusing on a simplified scenario in which analytic expressions can be obtained. Our conclusions are summarized in Sec.~\ref{sec:conclusion}, and additional technical details are provided in the Appendices.

\section{Model}\label{sec:model}
We consider a scale-invariant general two-Higgs-doublet model (SI-g2HDM), whose Lagrangian is obtained by removing all mass terms from the general 2HDM (g2HDM) (for an extensive review of the 2HDM, see Ref.~\cite{Branco:2011iw}; see also 
Refs.~\cite{Botella:1994cs,Davidson:2005cw,Gunion:2005ja,Haber:2010bw,Crivellin:2013wna} for the g2HDM):
\begin{align}
\mathcal{L}_{\text{SI-g2HDM}} & = \mathcal{L}_{\text{g2HDM}}|_{\text{w/o~mass terms}}.
\end{align}
In the so-called generic basis, both Higgs doublets can simultaneously acquire vacuum expectation values (VEVs). However, since the two doublets are not distinguished by definite quantum numbers, one may instead adopt the Higgs basis~\cite{Botella:1994cs} (also known as the Georgi basis~\cite{Georgi:1978ri}), in which only one doublet develops a VEV while the other does not. The two bases are physically equivalent, and the choice can be made according to convenience. In what follows, we work in the Higgs/Georgi basis, where the direction of spontaneous symmetry breaking (SSB) is more transparent than in the generic basis.

The tree-level scalar potential is purely quartic and takes the form
\begin{align}
V_0(H_{1}, H_2)
&=\frac{\Lambda_1}{2}(H_1^\dagger H_1)^2
	+\frac{\Lambda_{2}}{2}(H_2^\dagger H_2)^2 
	+\Lambda_{3}( H_1^\dagger H_1)(H_2^\dagger H_2)
	+\Lambda_{4}( H_1^\dagger H_2)(H_2^\dagger H_1) \nonumber \\
&\quad	
+\left[
	\frac{\Lambda_{5}}{2}(H_1^{\dagger}H_2)^{2}
	+\Big\{\Lambda_6( H_1^\dagger H_1)
	+\Lambda_7(H_2^\dagger H_2)
	\Big\}( H_1^\dagger H_2)
	+{\rm H.c.}
\right],
\end{align}
where $\Lambda_{1-4}$ are real parameters, whereas $\Lambda_{5-7}$ may in general be complex.
Although the individual phases are unphysical, the following rephasing-invariant combinations serve as physical CP-violating invariants~\cite{Botella:1994cs,Davidson:2005cw,Gunion:2005ja,Haber:2010bw}:  
\begin{align}
\mathrm{Im}(\Lambda_5^* \Lambda_6^2), \quad 
\mathrm{Im}(\Lambda_5^* \Lambda_7^2), \quad 
\mathrm{Im}(\Lambda_6 \Lambda_7^*).
\label{CPinvHiggs}
\end{align}
The Higgs doublets are parameterized as
\begin{equation}
H_1(x)=
\left(
\begin{array}{c}
G^+(x)\\
\frac{1}{\sqrt{2}}\Big(v+h_1'(x)+iG^0(x)\Big)
\end{array}
\right),\quad
H_2(x)=
\left(
\begin{array}{c}
H^+(x)\\
\frac{1}{\sqrt{2}}\Big(h_2'(x)+iA(x)\Big)
\end{array}
\right),
\end{equation}
where $v\simeq 246$ GeV, $G^{0}$ and $G^+$ [$G^-\equiv (G^+)^*$] are the Nambu-Goldstone (NG) bosons, and $H^+$ [$H^-\equiv (H^+)^*$] are the charged Higgs bosons.
The CP-even Higgs bosons $h_1'$ and $h_2'$ and the CP-odd Higgs boson $A$ would mix if CP is violated. 

In the Yukawa sector, the Lagrangian is given by 
\begin{align}
-\mathcal{L}_{\text{Yukawa}}&=\bar{q}_{iL}(Y_{0ij}^{d}H_{1}+\rho_{0ij}^{d}H_{2})d_{jR}
+\bar{q}_{iL}(Y_{0ij}^{u}\tilde{H}_{1}+\rho_{0ij}^{u}\tilde{H}_{2})u_{jR}\nonumber \\
&\quad +\bar{\ell}_{iL}(Y_{0ij}^{e}H_{1}+\rho_{0ij}^{e}H_{2})e_{jR}+\mbox{H.c.},
\label{Lag_Y}
\end{align}
where $i,j$ are generation indices, and $\tilde{H}_{1,2}=i\tau_{2}H_{1,2}^{*}$ with $\tau_2$ denoting the second Pauli matrix. 
One can diagonalize $Y_0^f$ by bi-unitary transformations, while $\rho_0^f$ remain the nondiagonal 3-by-3 complex matrices: 
\begin{align}
V_L^f Y_0^f V_R^f & = Y_\text{diag}^f, \quad V_L^f \rho_0^f V_R^f = \rho^f,
\label{biUnitary}
\end{align}
where each element of $Y_\text{diag}^f$ is the mass-giving Yukawa coupling, $y_f = \sqrt{2}m_f/v$.
The couplings $\rho^f$ can be CP-violating sources, and their off-diagonal elements induce tree-level flavor-changing neutral current processes, which are constrained by experimental data~\cite{Crivellin:2013wna}. As demonstrated in the ordinary massive g2HDM~\cite{Fuyuto:2019svr},  complex phases of $\rho^f$ could play a role in EWBG scenarios. 
The full expressions of the Yukawa interactions in the mass eigenbasis are shown in Appendix~\ref{app:Hcoup}.
Here, we present the Yukawa interactions in the \textit{nearly} CP-conserving limit, where the CP mixing is absent in the tree-level Higgs sector. This situation pertains to our case, as demonstrated in the following section.
The diagonalizing matrix for the neutral Higgs bosons and the mass eigenstates in the CP-conserving limit is defined as
\begin{align}
\left(
\begin{array}{c}
h_1'\\
h_2'\\
A
\end{array}
\right)
&=
\begin{pmatrix}
c_\gamma & s_\gamma & 0 \\
-s_\gamma & c_\gamma & 0 \\
0 & 0 & 1
\end{pmatrix}
\left(
\begin{array}{c}
H\\
h\\
A
\end{array}
\right),\label{diagonalization_tree}
\end{align}
where the $0\le \gamma\le \pi$. We define the SM-like limit  (also known as the \textit{alignment} limit) as $\gamma=\pi/2$, i.e., $s_\gamma=1$.
With this definition, the Yukawa interactions of the neutral Higgs bosons can be written as 
\begin{align}
-\mathcal{L}_{\rm Yukawa}
&\ni
\bar{f}_{iL}
\left[
	\frac{y_i^{f}}{\sqrt{2}}s_\gamma\delta_{ij}
	+\frac{\rho_{ij}^{f}}{\sqrt{2}}c_\gamma
\right]f_{jR} h
+\bar{f}_{iL}
\left[
	\frac{y_i^{f}}{\sqrt{2}}c_\gamma\delta_{ij}
	-\frac{\rho_{ij}^{f}}{\sqrt{2}}s_\gamma
\right]f_{jR} H \nonumber\\
&\quad+\frac{i}{\sqrt{2}}
\bigg[
	-\bar{u}_{iL}\rho_{ij}^uu_{jR}
	+\bar{d}_{iL}\rho_{ij}^{d}d_{jR} 
	+\bar{e}_{iL}\rho_{ij}^{e}e_{jR}
\bigg]A +{\rm H.c.}.
\end{align}

This implies that CP violation is transmitted to the $h$ sector through the mixing angle $\gamma$. In the SM-like limit, $h$ becomes a purely CP-even state, while the Yukawa couplings of $H$ are entirely governed by $\rho^f$ and are independent of the mass-generating Yukawa couplings $y^f$. 
By contrast, the Yukawa couplings of $A$ are consistently determined by $\rho^f$. 
These features stand in sharp contrast to those of the softly broken $Z_2$-symmetric 2HDMs. 
In this work, we restrict our attention to the diagonal elements of $\rho^f$ and neglect the off-diagonal ones. 
For notational simplicity, we denote the diagonal elements as $\rho_{tt} \equiv \rho_{33}^u$, $\rho_{bb} \equiv \rho_{33}^d$, and $\rho_{ee} \equiv \rho_{11}^e$, etc. 
In the presence of $\rho_{tt}$ and $\rho_{ee}$, which are the relevant couplings in the current analysis, the rephasing invariants, 
in addition to Eq.~(\ref{CPinvHiggs}), are given by~\cite{Botella:1994cs,Haber:2010bw}
\begin{align}
&\mathrm{Im}(\rho_{tt}^2 \Lambda_5), \quad \mathrm{Im}(\rho_{tt} \Lambda_6), \quad \mathrm{Im}(\rho_{tt} \Lambda_7), \quad \mathrm{Im}(\rho_{ee}^2 \Lambda_5^*), \quad \mathrm{Im}(\rho_{ee} \Lambda_6^*), \quad 
  \mathrm{Im}(\rho_{ee} \Lambda_7^*), \nonumber\\
&\mathrm{Im}(\rho_{tt} \rho_{ee}), \quad \mathrm{Im}(\rho_{tt} \rho_{ee}^* \Lambda_5).
\label{CPinvY}
\end{align}
Although the analysis can be carried out directly in terms of these invariants, it can be simplified by choosing a basis in which $\Lambda_5$ is real. We adopt this convention in the following discussion.

\subsection{Higgs sector at the tree level}\label{subsec:treeH}
The first derivatives of $V_0$ with respect to the neutral Higgs fields (called tadpole conditions, actually the flatness conditions at the tree level) are, respectively, given by\footnote{$T_{G^0}^{(0)} = \left\langle\partial V_0/\partial G^0\right\rangle = 0$ is guaranteed by the gauge invariance.}
\begin{align}
T_{h_1'}^{(0)} &= \left\langle\frac{\partial V_0}{\partial h_1'}\right\rangle = \frac{1}{2}\Lambda_1v^3 = 0,\label{dV0dh1p} \\
T_{h_2'}^{(0)} &= \left\langle\frac{\partial V_0}{\partial h_2'}\right\rangle = \frac{1}{2}\text{Re}\Lambda_6v^3 = 0,\label{dV0dh2p} \\
T_{A}^{(0)} &= \left\langle\frac{\partial V_0}{\partial A}\right\rangle = -\frac{1}{2}\text{Im}\Lambda_6v^3 = 0, \label{dV0dA}
\end{align}
where the bracket denotes that the fluctuation fields are taken to be zero after the derivatives.
The trivial solution $v=0$ defines the massless theory. For $v\neq0$, the solutions are $\Lambda_1=\text{Re}\Lambda_6=\text{Im}\Lambda_6 = 0$. However, the vacuum energy of $V_0$ turns out to be zero for any $v\neq0$, namely, $V_0$ is flat in the $h_1'$ direction, meaning that the EW symmetry is not spontaneously broken at the tree level.

Generally, the mass matrix of the neutral Higgs bosons takes the 4-by-4 form
\begin{equation}
\frac{1}{2}
\left(
\begin{array}{cccc}
h_1' & h_2' & G^0 & A
\end{array}
\right){\cal M}_0^2
\left(
\begin{array}{c}
h_1'\\
h_2'\\
G^0\\
A
\end{array}
\right),\quad {\cal M}_0^2=
\left(
\begin{array}{cc}
{\cal M}_S^2 & {\cal M}_{SP}^2\\
({\cal M}_{SP}^2)^T & {\cal M}_P^2
\end{array}
\right).
\end{equation}
Since the NG bosons are separated from the physical spectrum in the vacuum, the mass matrix is reduced to the 3-by-3 form
\begin{align}
\mathcal{M}_N^2 &= 
\begin{pmatrix}
\frac{3}{2}\Lambda_1v^2 & \frac{3}{2}\text{Re}\Lambda_6 v^2 & -\frac{3}{2}\text{Im}\Lambda_6 v^2 \\
\frac{3}{2}\text{Re}\Lambda_6 v^2 & \frac{1}{2}\Lambda_{345}v^2 & 0 \\
-\frac{3}{2}\text{Im}\Lambda_6 v^2 & 0 & \frac{1}{2}(\Lambda_3+\Lambda_4-\Lambda_5)v^2
\end{pmatrix} 
=
\begin{pmatrix}
0 & 0  & 0 \\
0 & \frac{1}{2}\Lambda_{345}v^2 & 0 \\
0 & 0 & \frac{1}{2}(\Lambda_3+\Lambda_4-\Lambda_5)v^2
\end{pmatrix}, 
\end{align}
where $\Lambda_{345} = \Lambda_3+ \Lambda_4+\Lambda_5$, and we have used the tadpole conditions Eqs.~(\ref{dV0dh1p})-(\ref{dV0dA}) in the second equality. 
Therefore, the CP mixing is absent at this order, and the Higgs boson masses are, respectively, given by
\begin{align}
m_h^2 = 0,\quad  m_H^2 = \frac{1}{2}\Lambda_{345} v^2, \quad m_A^2 =\frac{1}{2}(\Lambda_3+\Lambda_4-\Lambda_5)v^2.
\label{mhmH_tree}
\end{align}
The appearance of the massless particle is the consequence of the classical scale symmetry. 
Moreover, because of the classical symmetry, $h_1'$ does not mix with the additional Higgs bosons and becomes SM-like, which is favored by the current LHC data, and is also the consequence of this symmetry.

The mass matrix of the charged scalars is also derived from the second derivatives of $V_0$:
\begin{align}
\begin{pmatrix}
G^+ & H^+
\end{pmatrix}
\mathcal{M}_\pm^2 
\begin{pmatrix}
G^- \\
H^-
\end{pmatrix}
\end{align}
where
\begin{align}
\mathcal{M}_\pm^2 &= 
\begin{pmatrix}
\frac{1}{2}\Lambda_1v^2 & \frac{1}{2}(\text{Re}\Lambda_6-i\text{Im}\Lambda_6) v^2 \\
\frac{1}{2}(\text{Re}\Lambda_6+i\text{Im}\Lambda_6) v^2 & \frac{1}{2}\Lambda_3v^2
\end{pmatrix}
=
\begin{pmatrix}
0 & 0 \\
0 & \frac{1}{2}\Lambda_3v^2
\end{pmatrix}.
\end{align}
The flatness conditions (\ref{dV0dh1p}) - (\ref{dV0dA}) are used in the second equality. 
Thus, the charged Higgs boson mass at the tree level is  
\begin{align}
m_{H^\pm}^2 = \frac{1}{2}\Lambda_3v^2.
\label{mAmch_tree}
\end{align}
Note that $m_A^2 = m_{H^\pm}^2+\frac{1}{2}(\Lambda_4-\Lambda_5)v^2$. 
As is well known, a mass splitting between $m_A$ and $m_{H^\pm}$ signals custodial SU(2) symmetry breaking, which is tightly constrained by the $T$ parameter. Therefore, the choice $m_{H^\pm} \simeq m_A$ is favored to remain consistent with experimental data~\cite{ParticleDataGroup:2024cfk}. An alternative way to evade the $T$ parameter constraint is the so-called \emph{twisted custodially symmetry} scenario, where $m_{H^\pm} \simeq m_H$~\cite{Gerard:2007kn}. We focus exclusively on those two cases in this study. 

Here, let us see the tree-level relationships between the original parameters and Higgs masses. We have seven quartic couplings in $V_0$, $\Lambda_{1-7}$. 
The flatness conditions enforce that $\Lambda_1=\text{Re}\Lambda_6=\text{Im}\Lambda_6=0$, while $\Lambda_3$, $\Lambda_4$, and $\Lambda_5$ can be expressed by 
\begin{align}
\Lambda_3 & = \frac{2m_{H^\pm}^2}{v^2}, \label{L3_tree}\\
\Lambda_4 & = \frac{1}{v^2}(m_H^2+m_A^2-2m_{H^\pm}^2),  \label{L4_tree} \\
\Lambda_5 & = \frac{1}{v^2}(m_H^2-m_A^2).  \label{L5_tree}
\end{align}
In contrast, $\Lambda_{2}$, $\text{Re}\Lambda_7$, and $\text{Im}\Lambda_7$ are unrelated to the Higgs VEV and masses at the tree level. 
\subsection{Higgs sector at the one-loop level}\label{subsec:1LHiggs}
Now, we analyze the Higgs sector at the one-loop level. 
The tadpole conditions at this order are
\begin{align}
T_{h_1'} &= T_{h_1'}^{(0)}+T_{h_1'}^{(1)} = \frac{1}{2}\Lambda_1v^3+\left\langle \frac{\partial V_1}{\partial h_1'}\right\rangle = 0, \label{dVdh1p} \\
T_{h_2'} &= T_{h_2'}^{(0)}+T_{h_2'}^{(1)} = \frac{1}{2}\text{Re}\Lambda_6v^3+\left\langle \frac{\partial V_1}{\partial h_2'}\right\rangle = 0, \label{dVdh2p} \\
T_{A} &= T_{A}^{(0)}+T_{A}^{(1)} = -\frac{1}{2}\text{Im}\Lambda_6v^3+\left\langle \frac{\partial V_1}{\partial A}\right\rangle = 0, \label{dVdA}
\end{align}
where $V_1$ denotes the one-loop effective potential. In this work, we regularize it in the $\overline{\text{MS}}$ scheme. 
From the above tadpole conditions, one gets
\begin{align}
\Lambda_1 = -\frac{2}{v^3}\left\langle \frac{\partial V_1}{\partial h_1'}\right\rangle,\quad
\text{Re}\Lambda_6 = -\frac{2}{v^3}\left\langle \frac{\partial V_1}{\partial h_2'}\right\rangle,\quad
\text{Im}\Lambda_6 = \frac{2}{v^3}\left\langle \frac{\partial V_1}{\partial A}\right\rangle.
\label{tad1L}
\end{align}
 The one-loop mass matrix of the neutral Higgs is
\begin{align}
\mathcal{M}_N^2 =
\begin{pmatrix}
-\frac{3}{v}T_{h_1'}^{(1)}+\left\langle \frac{\partial^2 V_1}{\partial h_1'^2}\right\rangle & \frac{3}{v}T_{h_2'}^{(1)} +\left\langle \frac{\partial^2 V_1}{\partial h_1'\partial h_2'}\right\rangle & -\frac{3}{v}T_{A}^{(1)}+\left\langle \frac{\partial^2 V_1}{\partial h_1' \partial A}\right\rangle \\
\frac{3}{v}T_{h_2'}^{(1)}+\left\langle \frac{\partial^2 V_1}{\partial h_1'\partial h_2'}\right\rangle & \frac{1}{2}\Lambda_{345}v^2+\left\langle \frac{\partial^2 V_1}{\partial h_2'^2}\right\rangle & \left\langle \frac{\partial^2 V_1}{\partial  h_2' \partial A}\right\rangle \\
-\frac{3}{v}T_{A}^{(1)}+\left\langle \frac{\partial^2 V_1}{\partial h_1' \partial A}\right\rangle & \left\langle \frac{\partial^2 V_1}{\partial  h_2' \partial A}\right\rangle & \frac{1}{2}(\Lambda_3+\Lambda_4-\Lambda_5)v^2+\left\langle \frac{\partial^2 V_1}{\partial  A^2}\right\rangle
\end{pmatrix},
\end{align}
where we have used the one-loop tadpole conditions (\ref{dVdh1p})-(\ref{dVdA}).
The off-diagonal elements, including the CP mixings, are nonzero at the one-loop level. 
Consequently, $h_1'$ is neither massless nor a pure CP-even state. 

Following the Gildener-Weinberg (GW) method, we first find the flat direction of $V_0$ and investigate the SSB in that direction. 
The tadpole conditions while maintaining $\Lambda_1 =\text{Re}\Lambda_6=\text{Im}\Lambda_6=0$ at some scale (denoted as $\bar{\mu}_{\text{GW}}$) become
\begin{align}
T_{h_1'}^{(1)} = \left\langle \frac{\partial V_1}{\partial h_1'}\right\rangle=0,\quad
T_{h_2'}^{(1)} =  \left\langle \frac{\partial V_1}{\partial h_2'}\right\rangle=0,\quad
T_{A}^{(1)} = \left\langle \frac{\partial V_1}{\partial A}\right\rangle=0.
\end{align}
The first condition gives 
\begin{align}
T_{h_1'}^{(1)} & = \sum_{\substack{i=H,A,H^\pm,\\ W^\pm, Z, t,b}}n_i \frac{m_i^4}{16\pi^2 v}\left[\ln\frac{m_i^2}{\bar{\mu}_{\text{GW}}^2}-c_i+\frac{1}{2} \right] =
 \left[4A+2B+4B\ln\frac{v^2}{\bar{\mu}_\text{GW}^2}\right]v^3=0,
\end{align}
where $n_i$ denotes degrees of freedom with a statistical sign: $n_H = n_A = 1$, $n_{H^\pm} =2$, $n_W = 6$, $n_Z = 3$, $n_t=n_b = -4N_C^f$ with $N_C^{f}=3$ being the number of color, while $c_i=3/2$ for scalars and fermions and 5/6 for gauge bosons. All masses $m_i$ are defined at the tree level. The parameters $A$ and $B$ are, respectively, given by
\begin{align}
A &= \frac{1}{64\pi^2v^4}
\bigg[
	m_H^4\left(\ln\frac{m_H^2}{v^2}-\frac{3}{2}\right)
	+m_A^4\left(\ln\frac{m_A^2}{v^2}-\frac{3}{2}\right)
	+2m_{H^\pm}^4\left(\ln\frac{m_{H^\pm}^2}{v^2}-\frac{3}{2}\right)\nonumber\\
&\hspace{2cm}	
	+6m_W^4\left(\ln\frac{m_W^2}{v^2}-\frac{5}{6}\right)	
	+3m_Z^4\left(\ln\frac{m_Z^2}{v^2}-\frac{5}{6}\right) \nonumber\\
&\hspace{2cm}
	-4N_C^f
	\left\{
	m_t^4\left(\ln\frac{m_t^2}{v^2}-\frac{3}{2}\right)
	+m_b^4\left(\ln\frac{m_b^2}{v^2}-\frac{3}{2}\right)	
	\right\}	
\bigg], \\
B &= \frac{1}{64\pi^2 v^4}
\Big[
	m_H^4 + m_A^4+2m_{H^\pm}^4 + 6m_W^4+3m_Z^4-4N_C^f(m_t^4+m_b^4)
\Big].\label{B}
\end{align}
The vacuum energy is found to be $ V_1(v) = -Bv^4/2$.
Thus, the EW symmetry is spontaneously broken if $B>0$ and $v$ is generated by $\bar{\mu}_{\text{GW}}$ (dimensional transmutation):
\begin{align}
v^2 = \bar{\mu}_{\text{GW}}^2 e^{-1/2-A/B}.
\label{DT}
\end{align}
As seen from Eq.~(\ref{B}), $B$ would become negative if the additional Higgs bosons, $H$, $A$, and $H^\pm$ are absent. In other words, SSB does not occur in the massless SM. 
Furthermore, the magnitude of $\ln (v^2/\bar{\mu}_{\text{GW}}^2)$ is not sizable as the $A$ and $B$ are the same order in magnitude.

The one-loop tadpole conditions with respect to $h_2'$ are identified as
\begin{align} 
T_{h'_2}^{(1)} 
&= \frac{v}{16\pi^2}
\bigg[
	-2\sum_{f} N_C^fy_f\text{Re}\rho_{ff}m_f^2\left(\ln\frac{m_f^2}{\bar{\mu}_\text{GW}^2}-1\right) \nonumber \\
&\hspace{1.5cm}
	+\frac{1}{2}\text{Re}\Lambda_7
	\bigg\{
		3m_H^2\left(\ln\frac{m_H^2}{\bar{\mu}_\text{GW}^2}-1\right)+m_A^2\left(\ln\frac{m_A^2}{\bar{\mu}_\text{GW}^2}-1\right)
		+2m_{H^\pm}^2\left(\ln\frac{m_{H^\pm}^2}{\bar{\mu}_\text{GW}^2}-1\right)
	\bigg\}
\bigg],
\end{align}
where we take $s_\gamma=1$. The error in this approximation is a two-loop order, which has been discarded. 
Similarly, the one-loop tadpole condition with respect to $A$ is 
\begin{align}
T_{A}^{(1)} 
&= \frac{v}{16\pi^2}
\bigg[
	\pm 2\sum_{f}N_C^fy_f\text{Im}\rho_{ff}m_f^2\left(\ln\frac{m_f^2}{\bar{\mu}_\text{GW}^2}-1\right) \nonumber \\
&\hspace{1.5cm}
	+\frac{1}{2}\text{Im}\Lambda_7
	\bigg\{
		m_H^2\left(\ln\frac{m_H^2}{\bar{\mu}_\text{GW}^2}-1\right)+3m_A^2\left(\ln\frac{m_A^2}{\bar{\mu}_\text{GW}^2}-1\right)
		+2m_{H^\pm}^2\left(\ln\frac{m_{H^\pm}^2}{\bar{\mu}_\text{GW}^2}-1\right)
	\bigg\}
\bigg],
\end{align}
where the upper sign is for up-type fermions and the lower sign is for down-type fermions.

Solving $T_{h_2'}^{(1)}=0$ for $\text{Re}\Lambda_7$ and $T_{A}^{(1)}=0$ for $\text{Im}\Lambda_7$, one obtains
\begin{align}
\text{Re}\Lambda_7& = \frac{4\sum_fN_C^fy_f\text{Re}\rho_{ff}m_f^2\left(\ln\frac{m_f^2}{\bar{\mu}_\text{GW}^2}-1\right)}{3m_H^2\left(\ln\frac{m_H^2}{\bar{\mu}_\text{GW}^2}-1\right)+m_A^2\left(\ln\frac{m_A^2}{\bar{\mu}_\text{GW}^2}-1\right)
		+2m_{H^\pm}^2\left(\ln\frac{m_{H^\pm}^2}{\bar{\mu}_\text{GW}^2}-1\right)},\label{ReL7}\\
\text{Im}\Lambda_7 & = \frac{\pm 4\sum_fN_C^fy_f\text{Im}\rho_{ff}m_f^2\left(\ln\frac{m_f^2}{\bar{\mu}_\text{GW}^2}-1\right)}{m_H^2\left(\ln\frac{m_H^2}{\bar{\mu}_\text{GW}^2}-1\right)+3m_A^2\left(\ln\frac{m_A^2}{\bar{\mu}_\text{GW}^2}-1\right)
		+2m_{H^\pm}^2\left(\ln\frac{m_{H^\pm}^2}{\bar{\mu}_\text{GW}^2}-1\right)}.\label{ImL7}
\end{align}
Therefore, $\Lambda_7$ is fixed by $\rho_{ff}$ and additional Higgs bosons.

The (1,1)-element of $\mathcal{M}_N^2$ is given by
\begin{align}
(\mathcal{M}_N^2)_{11} & = \left\langle \frac{\partial^2 V_1}{\partial h_1'^2}\right\rangle = 8Bv^2,
\label{mhsq_MS}
\end{align}
where Eq.~(\ref{DT}) is used to eliminate $\bar{\mu}_{\text{GW}}$. Since the off-diagonal elements of $\mathcal{M}_N^2$ are one-loop suppressed, the approximate Higgs mass is given by $m_h^2\simeq 8Bv^2$. This implies that the requirement of $m_h\simeq 125$ GeV leads to a mass relation, $(m_H^4+m_A^4+2m_{H^\pm}^4)^{1/4}\simeq 540$ GeV. For a degenerate mass case $m_H=m_A=m_{H^\pm}\equiv m_\Phi$, one gets $m_\Phi\simeq 382$ GeV.
The sum rule for the mass spectrum does not significantly change even when the one-loop corrections are included. 
 
The (1,2) element of $\mathcal{M}_N^2$ is cast into the form 
\begin{align}
(\mathcal{M}_N^2)_{12}
&=\frac{-1}{16\pi^2}\bigg[
\sum_{f}4N_C^fy_f\text{Re}\rho_{ff}m_f^2\ln \frac{m_f^2}{\bar{\mu}^2_\text{GW}} \nonumber\\
&\hspace{1.5cm}
-\text{Re}\Lambda_7
\left(
	3m_H^2\ln \frac{m_H^2}{\bar{\mu}_\text{GW}^2}
	+m_A^2\ln \frac{m_A^2}{\bar{\mu}^2_\text{GW}}
	+2m_{H^\pm}^2\ln \frac{m_{H^\pm}^2}{\bar{\mu}_\text{GW}^2}
\right)
\bigg].
\end{align}
Note that $(\mathcal{M}_N^2)_{12}$ becomes zero for $\text{Re}\rho_{ff}=0$ because of vanishing $\text{Re}\Lambda_7$ as given by Eq.~(\ref{ReL7}).
On the other hand, the CP-mixing mass matrix elements are, respectively, given by
\begin{align}
(\mathcal{M}_N^2)_{13} 
& =\frac{1}{16\pi^2}
\bigg[
\mp 4N_C^fy_f\text{Im}\rho_{ff}m_f^2\ln\frac{m_f^2}{\bar{\mu}_\text{GW}^2} \nonumber\\
&\hspace{1.5cm}
	-\text{Im}\Lambda_7
	\left(
		m_H^2\ln\frac{m_H^2}{\bar{\mu}_\text{GW}^2}+3m_A^2\ln\frac{m_A^2}{\bar{\mu}_\text{GW}^2}+2m_{H^\pm}^2\ln\frac{m_{H^\pm}^2}{\bar{\mu}_\text{GW}^2}
	\right)
\bigg], \\
(\mathcal{M}_N^2)_{23}
& =\frac{1}{16\pi^2}
\bigg[
	\pm 4N_C^f\text{Re}\rho_{ff}\text{Im}\rho_{ff}m_f^2\ln\frac{m_f^2}{\bar{\mu}_\text{GW}^2}
\nonumber  \\
&\hspace{1.5cm}
	+\frac{v^2}{2}\text{Re}\Lambda_7\text{Im}\Lambda_7
	\bigg\{
		3\ln\frac{m_H^2}{\bar{\mu}_\text{GW}^2}+3\ln\frac{m_A^2}{\bar{\mu}_\text{GW}^2}+2\ln\frac{m_{H^\pm}^2}{\bar{\mu}_\text{GW}^2} \nonumber \\
&\hspace{4.5cm}
	+\frac{2}{m_H^2-m_A^2}\left[m_H^2\left(\ln\frac{m_H^2}{\bar{\mu}_\text{GW}^2}-1\right)
	-m_A^2\left(\ln\frac{m_A^2}{\bar{\mu}_\text{GW}^2}-1\right) \right]
	\bigg\}
\bigg].
\end{align}
Since both terms arise from the scalar-pseudoscalar mixings, they depend on the imaginary parts of the couplings, $\text{Im}\rho_{ff}$ and $\text{Im}\Lambda_7$. Moreover, these imaginary parts are related by the condition Eq.~(\ref{ImL7}), thus making the CP mixing zero for $\text{Im}\rho_{ff}=0$ only. Additinally, we note that $(\mathcal{M}_N^2)_{23}=0$ for $\text{Re}\rho_{ff}=0$.

The one-loop corrections to $(\mathcal{M}_N^2)_{22}$ and $(\mathcal{M}_N^2)_{33}$ are also included in our study, though their numerical impacts are subleading. For their explicit formulas, see Appendix~\ref{app:1LSigma}.

$\mathcal{M}_N^2$ is diagonalized by an orthogonal matrix $O$ as
\begin{align}
O^T\mathcal{M}_N^2 O = 
\begin{pmatrix}
m_{h_1}^2 & & \\
& m_{h_2}^2 & \\
& & m_{h_3}^2
\end{pmatrix},
\label{defO}
\end{align}
where $m_{h_1}\le m_{h_2}\le m_{h_3}$. We evaluate $O$ numerically. 
In this study, $h_1$ refers to the SM-like Higgs boson with a mass of $125$ GeV.
We will quantify the proximity of $m_{h_1}$ to the approximate Higgs mass $m_h=2\sqrt{2B}v$ below. 

The mass matrix of the charged scalars at the one-loop level is 
\begin{align}
\mathcal{M}_\pm^2 &= 
\begin{pmatrix}
\frac{1}{2}\Lambda_1v^2 + \left\langle \frac{\partial^2 V_1}{\partial G^+\partial G^-}\right\rangle & \frac{1}{2}(\text{Re}\Lambda_6-i\text{Im}\Lambda_6) v^2 + \left\langle \frac{\partial^2 V_1}{\partial G^+\partial H^-}\right\rangle\\
\frac{1}{2}(\text{Re}\Lambda_6+i\text{Im}\Lambda_6) v^2+\left\langle \frac{\partial^2 V_1}{\partial H^+\partial G^-}\right\rangle & \frac{1}{2}\Lambda_3v^2+\left\langle \frac{\partial^2 V_1}{\partial H^+\partial H^-}\right\rangle
\end{pmatrix} \nonumber \\
&=
\begin{pmatrix}
0 & 0 \\
0  & \frac{1}{2}\Lambda_3v^2+\left\langle \frac{\partial^2 V_1}{\partial H^+\partial H^-}\right\rangle
\end{pmatrix}
=
\begin{pmatrix}
0 & 0 \\
0  & (m_{H^\pm}^\mathrm{1\mbox{-}loop})^2
\end{pmatrix},
\label{mch_1L}
\end{align}
where the tadpole conditions are used in the second equality. The explicit formulas of the one-loop contribution to $m_{H^\pm}$ are displayed in Appendix.~\ref{app:1LSigma}.

We parametrize the Higgs couplings to the fermions as
\begin{align}
\mathcal{L}_{h_i\bar{f}f} = -h_i\bar{f}\left(g_{h_i\bar{f}f}^S+i\gamma_5g_{h_i\bar{f}f}^P\right)f, 
\end{align}
where
\begin{align}
g^S_{h_i\bar{f}f} &= \frac{1}{\sqrt{2}}\left[y_fO_{1i}+\text{Re}\rho_{ff}O_{2i}\pm\text{Im}\rho_{ff}O_{3i} \right], 
\label{gS}\\
g^P_{h_i\bar{f}f} &= \frac{1}{\sqrt{2}}\left[\text{Im}\rho_{ff}O_{2i}\mp \text{Re}\rho_{ff}O_{3i} \right],  
\label{gP}
\end{align}
where the upper signs for $f=u$ and lower signs for $f=d,e$.

The Higgs couplings to the gauge bosons are the same as in the ordinary 2HDM and are given by
\begin{align}
\mathcal{L}_{h_iVV} & =  \frac{1}{v}\sum_{i=1}^3g_{h_iVV}^{}h_i(m_Z^2Z_\mu Z^\mu+2m_W^2W_\mu^+W^{-\mu}), \\
\mathcal{L}_{h_ih_jZ} & = -\frac{g_2}{4c_W}\sum_{i,j=1}^3 g_{h_ih_jZ}^{}(h_i \bpartial_\mu h_j)Z^\mu, \\
\mathcal{L}_{h_iH^\pm W^\mp} & = -\frac{g_2}{2}\sum_{i=1}^3g_{h_iH^+W^-}(h_i i\bpartial_\mu H^+)W^{-\mu}+\mathrm{H.c.},
\end{align}
where $g_2$ denotes the SU(2$)_L$ gauge coupling, $\bpartial_\mu =\rpartial_\mu-\lpartial_\mu$, and 
\begin{align}
g_{h_iVV}^{} & = O_{1i}, \\
g_{h_ih_jZ}^{} & = O_{2i}O_{3j}-O_{2j}O_{3i}=\epsilon^{ijk}g_{h_kVV}, \label{ghihjZ} \\
g_{h_iH^+W^-} & = O_{2i}-iO_{3i},
\end{align}
where $\epsilon^{ijk}$ denotes the Levi-Civita symbol, and $\det(O)=1$ is assumed. 
It follows that $\sum_{i=1}^3 g_{h_iVV}^2 = 1$, and, for each $i$, $g_{h_iVV}^2 + |g_{h_iH^+W^-}|^2 = 1$. 
In the CP-conserving limit, these relations yield $g_{hVV}^{} = s_\gamma$, $g_{HVV}^{} = c_\gamma$, and $g_{AVV}^{} = 0$.

\begin{figure}[t]
\center
\includegraphics[width=7.5cm]{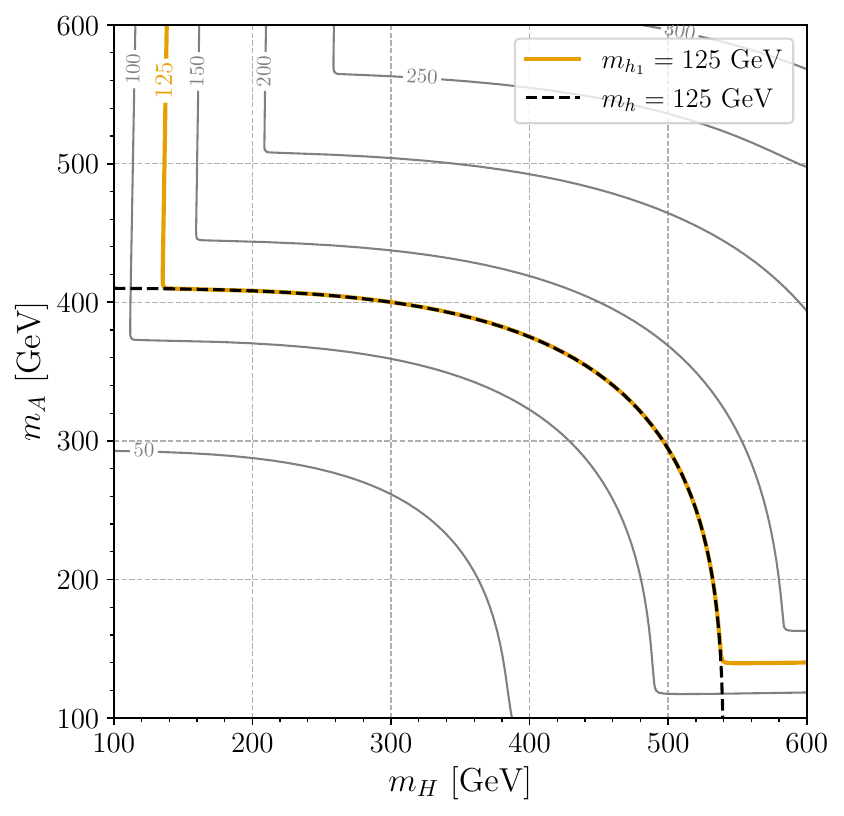}\hspace{1cm}
\includegraphics[width=7.5cm]{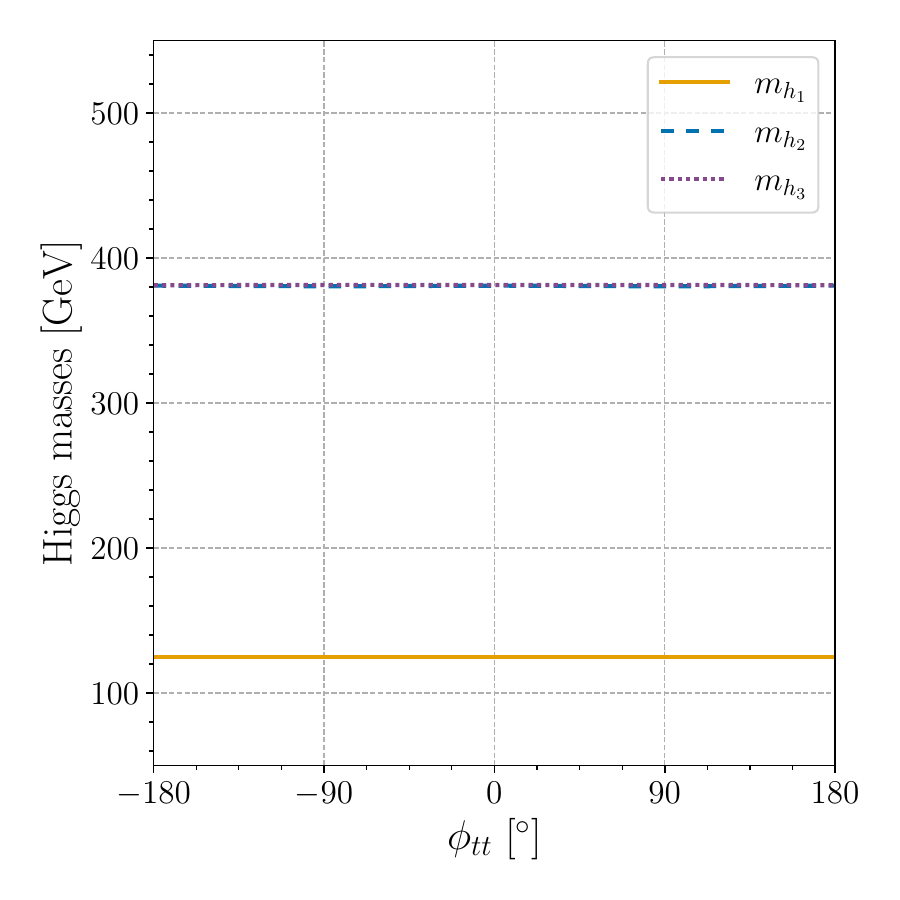}
\caption{(Left) The contours of $m_{h_1}$ in the $(m_H,m_A)$ plane, where the orange solid line represents $m_{h_1}=125$ GeV. For reference,  $m_h=2\sqrt{2B}v$ is also shown by the black dashed curve. We take $m_{H^\pm}=m_A$, $\Lambda_2=1.0$, and $|\rho_{tt}|=0.1$ and $\phi_{tt}=75^\circ$. (Right) The Higgs masses against $\phi_{tt}$ with fixed $m_A=m_{H^\pm}=382$ GeV, $|\rho_{tt}|=0.1$, and $\Lambda_2=1.0$. $m_H$ is numerically tuned to realize $m_{h_1}=125.0$ GeV. 
}
\label{fig:mh1_mH_mA}
\end{figure}

Here, let us summarize our input parameters. 
We have 5 real and 2 complex parameters in the scalar potential and extra complex Yukawa couplings:
\begin{align}
\Lambda_1,\quad \Lambda_2,\quad \Lambda_3,\quad \Lambda_4,\quad \Lambda_5, \quad \mathrm{Re}\Lambda_6,\quad  \mathrm{Im}\Lambda_6,\quad \mathrm{Re}\Lambda_7,\quad  \mathrm{Im}\Lambda_7; \quad \mathrm{Re}\rho_{ff},\quad \mathrm{Im}\rho_{ff}.
\end{align}
From the flatness condition, $\Lambda_1=\mathrm{Re}\Lambda_6=\mathrm{Im}\Lambda_6=0$. $\Lambda_3$, $\Lambda_3$, and $\Lambda_5$ are replaced by $m_H$, $m_A$, and $m_{H^\pm}$ using Eqs.~(\ref{L3_tree}), (\ref{L4_tree}), and (\ref{L5_tree}), where 
the heavy Higgs masses must satisfy $(m_H^4+m_A^4+2m_{H^\pm}^4)^{1/4}\simeq 540$ GeV.
Moreover, as noted in Sec.~\ref{subsec:treeH}, we consider the custodially symmetric ($m_A = m_{H^\pm}$) and twisted custodially symmetric ($m_H = m_{H^\pm}$) cases in order to avoid significant corrections to the $T$ parameter~\cite{Bertolini:1985ia}, thereby 
effectively reducing the number of independent heavy Higgs mass scales to one.\footnote{Since $m_A$ and $m_H$ are mixed by the CP violation, those symmetries are approximate.}
As we will discuss in detail in Sec.~\ref{sec:EDMs}, the region in which the electron EDM cancels depends on whether the Higgs spectrum is custodially symmetric or twisted custodially symmetric. 
Although one could in principle take \( (m_{h_2}, m_{h_3}, m_{H^\pm}^{\text{1-loop}}) \) as inputs, using \( (m_H, m_A, m_{H^\pm}) \) is technically simpler and makes the specification of the (twisted) custodial-symmetric Higgs spectrum more transparent.

$\mathrm{Re}\Lambda_7$ and $\mathrm{Im}\Lambda_7$ are replaced by Eqs.~(\ref{ReL7}) and (\ref{ImL7}).
$\Lambda_2$ and $\rho_{ff}$ are given as inputs. 
For late use, we parametrize $\rho_{ff}=|\rho_{ff}|e^{i\phi_{ff}}$.
Until we discuss the electron EDM in Sec.~\ref{sec:EDMs}, we assume that $\rho_{tt}\neq 0$ and $\rho_{ii}=0$ for all $i\neq t$.
In the remainder of this section, we mainly focus on the impacts of $\rho_{tt}$ on the Higgs mass spectrum and the Higgs couplings to gauge bosons and the top quark, and the combined analysis with the current electron EDM constraints will be presented in Sec.~\ref{sec:EDMs}.

In the left panel of Fig.~\ref{fig:mh1_mH_mA}, the contours of $m_{h_1}$ are shown as a function of $m_H$ and $m_A \,(=m_{H^\pm})$. 
As an illustrative example, we take $\Lambda_2 = 1.0$, $|\rho_{tt}| = 0.1$, and $\phi_{tt} = 75^\circ$. 
The orange solid line corresponds to $m_{h_1} = 125$ GeV. 
For comparison, the contour of $m_h = 2\sqrt{2B}\,v = 125$ GeV, which corresponds to the square root of $(\mathcal{M}_N^2)_{11}$, is also shown as a black dashed line. 
Except in the region where either $m_H$ or $m_A$ is less than about $140~\text{GeV}$, 
$m_{h_1}$ and $m_h$ are in good agreement with each other. 
In the parameter space where $m_{h_1} = 125~\text{GeV}$, we find $|(m_{h_1} - m_h)/m_{h_1}| \leq 1.3 \times 10^{-2}$. 

In the right panel, $m_{h_1}$ (solid, orange), $m_{h_2}$ (dashed, dark blue), and $m_{h_3}$ (dotted, purple) are shown for $m_A = m_{H^\pm} = 382~\text{GeV}$, $\Lambda_2 = 1.0$, and $|\rho_{tt}| = 0.1$, with $m_H$ numerically adjusted to yield 
$m_{h_1} = 125.0~\text{GeV}$. 
The resulting heavier Higgs boson masses are $m_{h_2} \in [380.5, 381.0]~\text{GeV}$ and $m_{h_3} \in [381.2, 381.5]~\text{GeV}$, 
respectively. This demonstrates that the tree-level Higgs masses provide a good approximation to the one-loop corrected values. The small mass splitting between $m_{h_3} (\simeq m_A)$ and $m_{H^\pm}$ remains, ensuring that the correction to the $T$ parameter is kept small. Nevertheless, the correction to the $S$ parameter is not always small to avoid the experimental constraint. 
Experimental consideration including the electron EDM will be given in Sec.~\ref{sec:EDMs}.

\begin{figure}[t]
\center
\includegraphics[width=7.5cm]{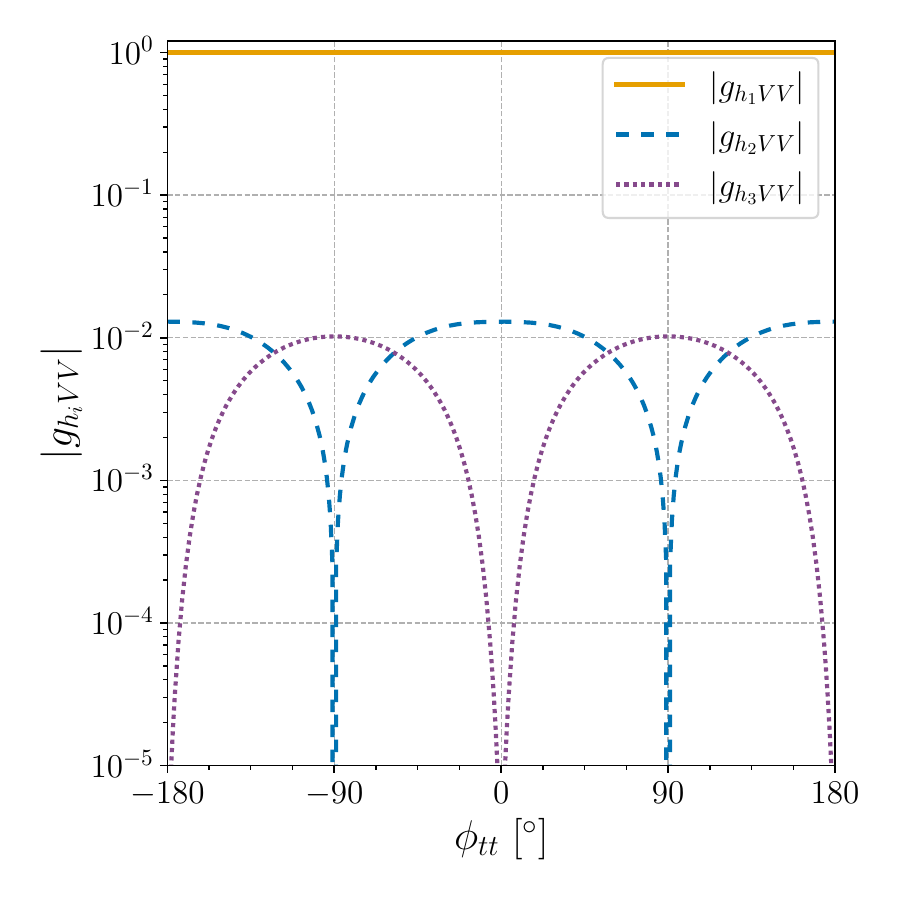} 
\includegraphics[width=7.5cm]{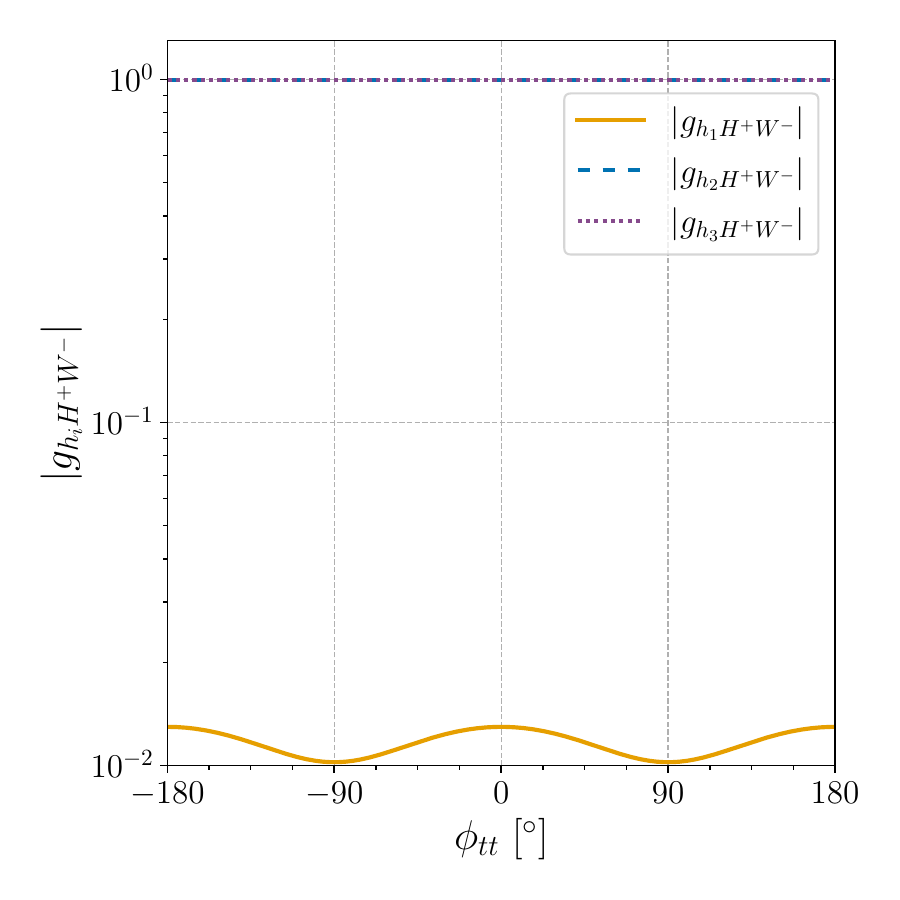} 
\caption{$|g_{h_iVV}|$ (left) and $|g_{h_iH^+W^-}|$ (right) as functions of $\phi_{tt}$. The input parameter is the same as in the right panel of Fig.~\ref{fig:mh1_mH_mA}}
\label{fig:Hmass_ghVV_ptt_mPhi382}
\end{figure}
In Fig.~\ref{fig:Hmass_ghVV_ptt_mPhi382}, the Higgs masses and their couplings to the gauge bosons are shown as functions of $\phi_{tt}$. The left panel displays $|g_{h_1VV}|$ (solid, orange), $|g_{h_2VV}|$ (dashed, dark blue), and $|g_{h_3VV}|$ (dotted, purple), while the right panel shows $|g_{h_1H^+W^-}|$ (solid, orange), $|g_{h_2H^+W^-}|$ (dashed, dark blue), and $|g_{h_3H^+W^-}|$ (dotted, purple). 
We observe that $|g_{h_1VV}|\simeq 1$ as a consequence of the one-loop suppressed off-diagonal elements $(\mathcal{M}_N^2)_{12}$ and $(\mathcal{M}_N^2)_{13}$, which is fully consistent with current LHC data. In contrast, $|g_{h_2VV}|$ and $|g_{h_3VV}|$ remain at most at the $\mathcal{O}(10^{-2})$ level, in accordance with the sum rule $\sum_{i=1}^3 g_{h_iVV}^2 = 1$. 
The vanishing of $|g_{h_2VV}|=|O_{12}|$ at $\phi_{tt} = \pm 90^\circ$ is explained by the zero of $(\mathcal{M}_N^2)_{12}\propto \cos\phi_{tt}$. From the right panel, we find $|g_{h_1H^+W^-}| = \mathcal{O}(10^{-2})$, while $|g_{h_2H^+W^-}|\simeq 1$ and $|g_{h_3H^+W^-}|\simeq 1$. This behavior follows from the sum rule $g_{h_iVV}^2 + |g_{h_iH^+W^-}|^2 = 1$ for each $i$. 
It is worth noting that the numerical values of the $g_{h_i h_j Z}$ couplings can be obtained from the relation $g_{h_i h_j Z} = \epsilon^{ijk} g_{h_k VV}$, as shown in Eq.~(\ref{ghihjZ}). 
We find $|g_{h_2 h_3 Z}| \simeq 1$, $|g_{h_1 h_2 Z}| \lesssim 10^{-2}$, and $|g_{h_1 h_3 Z}| \lesssim 10^{-2}$, respectively.

\begin{figure}[t]
\center
\includegraphics[width=7.5cm]{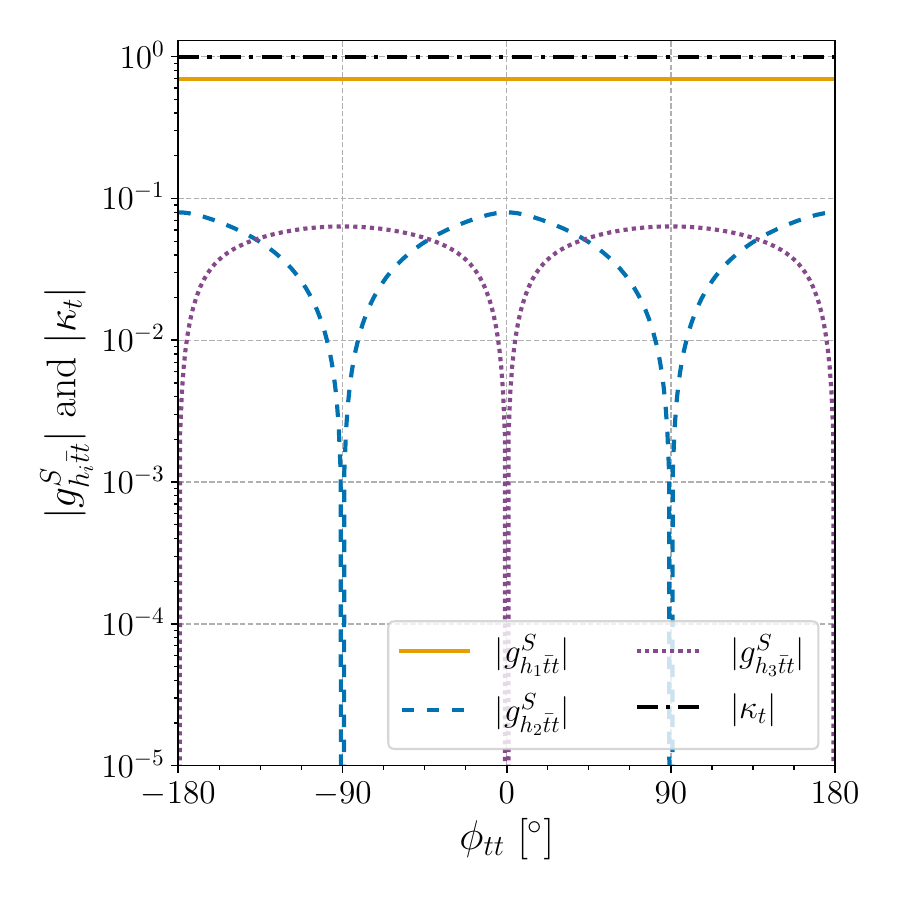} 
\includegraphics[width=7.5cm]{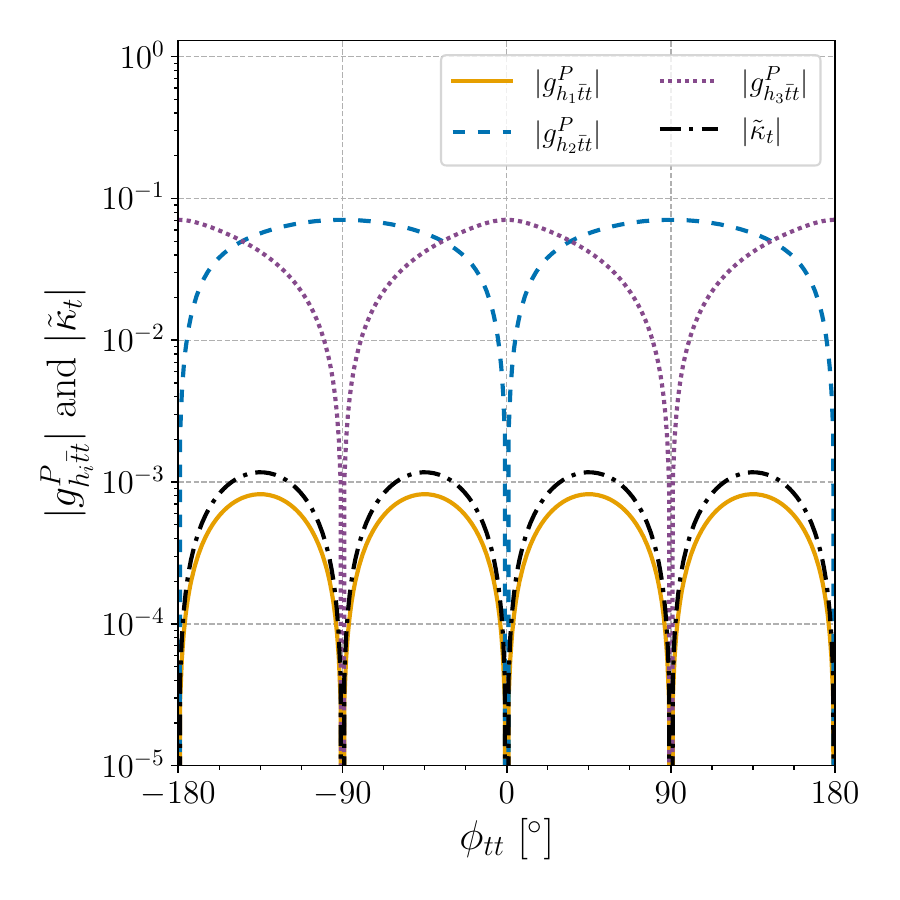} 
\caption{$|g_{h_i\bar{t}t}^S|$ (left) and $|g_{h_i\bar{t}t}^P|$ (right) as functions of $\phi_{tt}$. The input parameter is the same as in the right panel of Fig.~\ref{fig:mh1_mH_mA}
}
\label{fig:ghtt_ptt_mPhi382}
\end{figure}

In Fig.~\ref{fig:ghtt_ptt_mPhi382}, the Higgs couplings to the top quark are shown as functions of $\phi_{tt}$. In the left panel, 
$|g_{h_1\bar{t}t}^S|$ (solid, orange), $|g_{h_2\bar{t}t}^S|$ (dashed, dark blue), $|g_{h_3\bar{t}t}^S|$ (dotted, purple), 
and $\kappa_t$ (dot-dashed, black) are displayed, where $|\kappa_t| = v|g_{h_1\bar{t}t}^S|/m_t$. 
We find that $|g_{h_1\bar{t}t}^S|\simeq 0.7$, while $|g_{h_2\bar{t}t}^S|\lesssim 0.1$ and $|g_{h_3\bar{t}t}^S|\lesssim 0.1$. 
In the right panel, $|g_{h_i\bar{t}t}^P|$ and $\tilde{\kappa}_t$ are shown, where $|\tilde{\kappa}_t| = v|g_{h_1\bar{t}t}^P|/m_t$, with the line and color assignments corresponding to those in the scalar case. 
In contrast to $|g_{h_1\bar{t}t}^S|$, we find $|g_{h_1\bar{t}t}^P| < 10^{-3}$, while $|g_{h_2\bar{t}t}^P|\simeq |g_{h_3\bar{t}t}^S|$ and $|g_{h_3\bar{t}t}^P|\simeq |g_{h_2\bar{t}t}^S|$.

The CMS collaboration has placed limits on $\kappa_t$ and $\tilde{\kappa}_t$. 
The one-dimensional 95\% confidence intervals are given by~\cite{CMS:2022dbt}
\begin{align}
0.86 \leq \kappa_t \leq 1.26, 
\qquad 
-1.07 \leq \tilde{\kappa}_t \leq 1.07.
\end{align}
As shown in Fig.~\ref{fig:ghtt_ptt_mPhi382}, our results are fully consistent with 
these bounds. A parameter characterizing the CP nature of the $t\bar{t}h$ interaction, 
defined as a signal strength multiplier for the SM $t\bar{t}h$ cross section, is
\begin{align}
|f_\mathrm{CP}^{htt}| = \frac{|\tilde{\kappa}_t|^2}{|\kappa_t|^2+|\tilde{\kappa}_t|^2}.
\end{align}
From Fig.~\ref{fig:ghtt_ptt_mPhi382}, we obtain 
$|f_\mathrm{CP}^{htt}| < 1.4 \times 10^{-6}$, which is far below the current bound 
$|f_\mathrm{CP}^{htt}| < 0.55$ at 68\% CL~\cite{CMS:2022dbt}.

\section{Electric dipole moments}\label{sec:EDMs}
Here, we present the impacts of CP violation on electron EDM in this model. 
The electric dipole moment of a fermion $d_f$ is defined as the coefficient of the following operator
\begin{align}
\mathcal{L}_{\text{EDM}} = -\frac{i}{2}d_fF^{\mu\nu}\bar{f}\sigma_{\mu\nu}\gamma_5f,
\end{align}
where $F^{\mu\nu}$ denotes the electromagnetic field strength tensor. 
Currently, $d_e$ is constrained by the ACME and JILA experiments as~\cite{ACME:2018yjb,Roussy:2022cmp}
\begin{align}
|d_e^{\textrm{ACME}}| &< 1.1 \times 10^{-29}~e~\text{cm}~(90\%~\textrm{CL})\label{de_ACME18}, \\
|d_e^{\textrm{JILA}}| &< 4.1 \times 10^{-30}~e~\text{cm}~(90\%~\textrm{CL}).\label{de_JILA22}
\end{align}
In this work, we express the electron EDM as
\begin{align}
d_e = d_e^\mathrm{BZ} + d_e^\mathrm{kite},
\end{align}
where the first term arises from the Barr-Zee diagrams~\cite{Barr:1990vd,Ellis:2008zy,Cheung:2009fc,West:1993tk,Chang:2005ac,Ellis:2010xm,Cheung:2014oaa,Inoue:2014nva,Bowser-Chao:1997kjp,Abe:2013qla,Jung:2013hka,Hou:2021zqq,Hou:2023kho,Altmannshofer:2025nsl,Davila:2025goc}, 
while the second originates from the kite diagrams~\cite{Altmannshofer:2015qra,Altmannshofer:2025nsl,Altmannshofer:2020shb} at the two-loop level. 
We first analyze the Barr-Zee contributions to determine which model parameters control them, and subsequently consider the kite contributions.

\begin{figure}[t]
\center
\includegraphics[width=5cm]{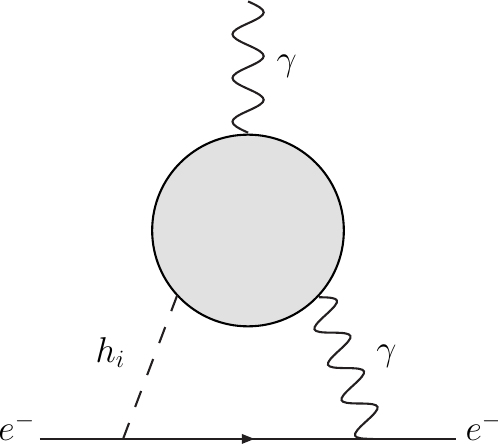}
\hspace{0.3cm}
\includegraphics[width=5cm]{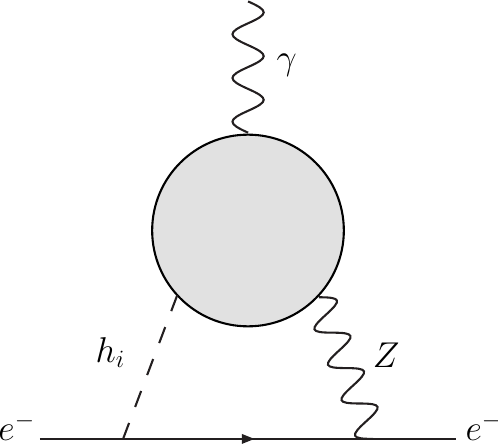}
\hspace{0.3cm}
\includegraphics[width=5cm]{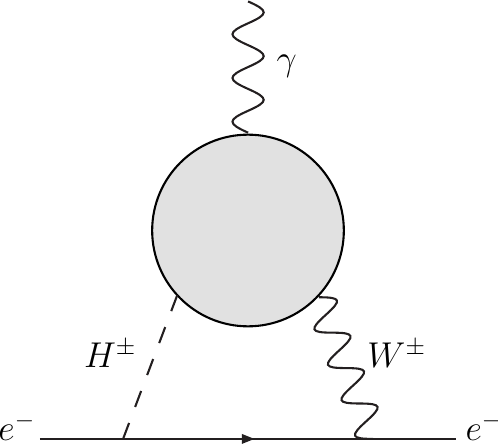}
\caption{Barr-Zee diagrams contributing to the electron EDM, which are, respectively, denoted as $d_{e}^{h\gamma}$, $d_{e}^{h Z}$, and $d_e^{HW}$.}
\label{fig:BZ}
\end{figure}

Depending on the particles running in the loop, we decompose the Barr-Zee contributions into three parts
\begin{align}
d_e^\mathrm{BZ} = d_e^{h \gamma}+d_e^{h Z}+d_e^{HW}.
\label{de_decomposed}
\end{align}
The Feynman diagram for each contribution is shown in Fig.~\ref{fig:BZ}. The shaded regions together represent the loops of the fermions, $W$ bosons, and charged Higgs bosons. 
Dominant contributions to $d_e$ usually come from $d_e^{h\gamma}$ among others, so we first focus on it.
For clarification, $d_e^{h\gamma}$ is further decomposed into 
\begin{align}
d_e^{h\gamma} = (d_e^{h \gamma})_{t}+(d_e^{h \gamma})_{W}+(d_e^{h \gamma})_{H^\pm},
\end{align}
where the subscripts denote the particles running the upper loop in the Barr-Zee diagrams, and each contribution is, respectively, given by~\cite{Barr:1990vd,Ellis:2008zy,Abe:2013qla}
\begin{align}
\frac{(d_{e}^{h\gamma})_t}{e}
&= \sum_{i=1}^3\frac{\alpha_{\rm em}}{6\pi^3m_t}
\left[
	g_{h_i\bar{e}e}^Pg_{h_i\bar{t}t}^Sf(\tau_{th_i})
	+g_{h_i\bar{e}e}^Sg_{h_i\bar{t}t}^Pg(\tau_{th_i})
\right]\equiv \frac{\alpha_{\rm em}}{6\pi^3m_t} \mathcal{X}_t^{h\gamma}, \\
\frac{(d_{e}^{h\gamma})_W}{e}
& = -\sum_{i=1}^3-\frac{\sqrt{2}G_F\alpha_\mathrm{em}v}{32\pi^3}g_{h_i\bar{e}e}^Pg_{h_iVV}\mathcal{J}_W^\gamma(m_{h_i})
\equiv -\frac{\sqrt{2}G_F\alpha_\mathrm{em}v}{32\pi^3}\mathcal{X}_W^{h\gamma}, \\
\frac{(d_{e}^{h\gamma})_{H^\pm}}{e}
&= -\sum_{i=1}^3 \frac{\alpha_{\text{em}}v}{32\pi^3m_{H^\pm}^2}
	\bar{\lambda}_{h_i H^+H^-}g_{h_i\bar{e}e}^P\mathcal{I}(\tau_{H^\pm h_i})
\equiv -\frac{\alpha_{\text{em}}v}{32\pi^3m_{H^\pm}^2}\mathcal{X}_{H^\pm}^{h\gamma},
\end{align}
where $\alpha_{\text{em}} = e^2 / 4\pi$ with $e = |e|$, $G_F = 1/\sqrt{2}v^2$,$f(\tau_{t h_i})$, $g(\tau_{t h_i})$, $\mathcal{J}_W^\gamma(m_{h_i})$, and $\mathcal{I}(\tau_{H^\pm h_i}) = f(\tau_{H^\pm h_i}) - g(\tau_{H^\pm h_i})$ with $\tau_{AB} = m_A^2 / m_B^2$ are loop functions defined in Appendix~\ref{app:edms}. It should be noted that the charged Higgs mass used in the EDM calculation is taken at the one-loop level. 

As studied in Refs.~\cite{Fuyuto:2019svr}, $\text{Im}\rho_{tt}$ could play a pivotal role in generating BAU in the g2HDM, while $\text{Im}\rho_{ee}$ can invoke a cancellation mechanism for the electron EDM in the presence of the nonzero $\rho_{tt}$~\cite{Fuyuto:2019svr}. We thus concentrate exclusively on those couplings and take other extra Yukawa couplings $\rho_{ff}$ to vanishingly small values in this work. 
In the following discussion, we consider two cases: 
\begin{itemize}
\item[(i)] $\rho_{tt}\neq0$ and $\rho_{ee}=0$.
\item[(ii)] $\rho_{tt}\neq0$ and $\rho_{ee}\neq0$.
\end{itemize}
Before proceeding to the details, we summarize the CP-violating phases considered in this study. 
Recalling the flatness conditions, $\Lambda_6 = 0$, $\mathrm{Re}\Lambda_7 \propto \mathrm{Re}\rho_{tt}$, and 
$\mathrm{Im}\Lambda_7 \propto \mathrm{Im}\rho_{tt}$, 
the rephasing invariants in Eq.~(\ref{CPinvY}) can be expressed as
\begin{align}
\mathrm{Im}\rho_{tt}^2, \quad \mathrm{Im}\,\rho_{ee}^2, \quad \mathrm{Im}(\rho_{tt}\rho_{ee}), \quad \mathrm{Im}(\rho_{tt}\rho_{ee}^*).
\end{align}
Thus, the electron EDM in this study is a function of these imaginary parts.
In Case (i), $d_e \propto \mathrm{Im}\rho_{tt}^2$. 
The vanishing of the EDM in the presence of an $\mathcal{O}(1)$ CP-violating phase should depend sensitively on the Higgs mass spectrum. 
In light of the heavy Higgs mass relation $(m_H^4 + m_A^4 + 2m_{H^\pm}^4)^{1/4} \simeq 540~\text{GeV}$, 
the solution space for $d_e=0$ is highly nontrivial. In Case (ii), $\rho_{ee}$ can contribute to realizing $d_e=0$, 
as in the ordinary g2HDM. However, the cancellation region is also constrained by the heavy Higgs mass relation.
Note that in the analysis of Ref.~\cite{Fuyuto:2019svr}, the kite contributions were not taken into account. 
Clarifying their numerical impact is therefore also one of the objectives of the present work.

\subsection{Case (i) $\rho_{tt}\neq0$ and $\rho_{ee}=0$}
In this case, the contribution arises solely from $(d_e^{h\gamma})_t$. 
To elucidate the dependence of the model parameters on the condition $(d_e^{h\gamma})_t = 0$, we consider a simplified example. Let us parametrize $\mathcal{M}_N^2$ and $O$ as
\begin{align}
\mathcal{M}_N^2 =
\begin{pmatrix}
m_h^2 & a\mathrm{Re}\rho_{tt} & b\mathrm{Im}\rho_{tt} \\
a\mathrm{Re}\rho_{tt} & m_H^2 &c\mathrm{Re}\rho_{tt}\mathrm{Im}\rho_{tt} \\
b\mathrm{Im}\rho_{tt} & c\mathrm{Re}\rho_{tt}\mathrm{Im}\rho_{tt} & m_A^2 
\end{pmatrix},\quad
O & = 
\begin{pmatrix}
1 & 0 & 0 \\
0 & c_3 & s_3 \\
0 & -s_3 & c_3
\end{pmatrix}
\begin{pmatrix}
c_2 & 0 & s_2  \\
0 & 1 & 0 \\
-s_2 & 0 & c_2
\end{pmatrix}
\begin{pmatrix}
c_1 & s_1 & 0 \\
-s_1 & c_1 & 0 \\
0 & 0 & 1
\end{pmatrix},
\end{align}
where $s_i=\sin\alpha_i$ and $c_i=\cos\alpha_i~(i=1,2,3)$. Note that $\alpha_1=\gamma$ defined in Eq.~(\ref{diagonalization_tree}).
Under the condition that $|a|$, $|b|$, and $|c|$ are smaller than the digonal elements and $m_{h}\neq m_H\neq m_A$, one would have $|\delta|\ll1$ and $|\alpha_{2,3}|\ll1$, where  $\delta =\pi/2- \alpha_1$, and thus the rotation matrix $O$ can be approximated as
\begin{align}
O& \simeq
\begin{pmatrix}
\hat{a}\mathrm{Re}\rho_{tt} & 1 & \hat{b}\mathrm{Im}\rho_{tt} \\
-1 & \hat{a}\mathrm{Re}\rho_{tt} & \hat{c}\mathrm{Re}\rho_{tt}\mathrm{Im}\rho_{tt} \\
\hat{c}\mathrm{Re}\rho_{tt}\mathrm{Im}\rho_{tt} & -\hat{b}\mathrm{Im}\rho_{tt}  & 1 
\end{pmatrix}+\mathcal{O}((\hat{a},\hat{b},\hat{c})^2),
\end{align}
where $\hat{a} = a/(m_h^2-m_H^2)$, $\hat{b} = b/(m_A^2-m_h^2)$, $\hat{c} = c/(m_A^2-m_H^2)$, and $\mathcal{O}((\hat{a},\hat{b},\hat{c})^2)$ denotes terms that are second order in $\hat{a}$, $\hat{b}$, and $\hat{c}$, including both diagonal (e.g., $\hat{a}^2$) and mixed terms (e.g., $\hat{a}\hat{b}$).
The factor $\mathcal{X}_t^{h\gamma}$ in the case of $\rho_{ee}=0$ is calculated as
\begin{align}
\mathcal{X}_t^{h\gamma} & = \frac{y_e}{2}\sum_iO_{1i}\Big[\text{Im}\rho_{tt}O_{2i}-\text{Re}\rho_{tt}O_{3i}\Big]g(\tau_{th_i}) \nonumber\\
&\simeq \frac{y_e}{4}|\rho_{tt}|^2\sin2\phi_{tt}
\Big[
	-\hat{a}\Delta g_{12}+\hat{b}\Delta g_{23}
\Big]+\mathcal{O}((\hat{a},\hat{b},\hat{c})^2),
\end{align}
where $\Delta g_{ij}=g(\tau_{th_i})-g(\tau_{th_j}).$
Therefore, the electron EDM is proportional solely to $\sin 2\phi_{tt}$, as required. 
This stands in contrast to the case discussed in Ref.~\cite{Fuyuto:2019svr}, where $(d_e^{h\gamma})_t \propto \sin \phi_{tt}$ in the nearly CP-conserving limit, in which CP violation is absent in the Higgs potential. 
It should be emphasized that $\mathcal{X}_t^{h\gamma}$ can vanish if the two terms in the parentheses cancel each other, corresponding to a specific Higgs mass spectrum. Moreover, the electron EDM also vanishes in the case of complete mass degeneracy, $m_{h_1} = m_{h_2} = m_{h_3}$, although such a situation cannot be realized in the 
present model. For related discussions of vanishing electron EDM scenarios, see, e.g., Ref.~\cite{Idegawa:2023bkh}.

\begin{figure}[t]
\center
\includegraphics[width=10cm]{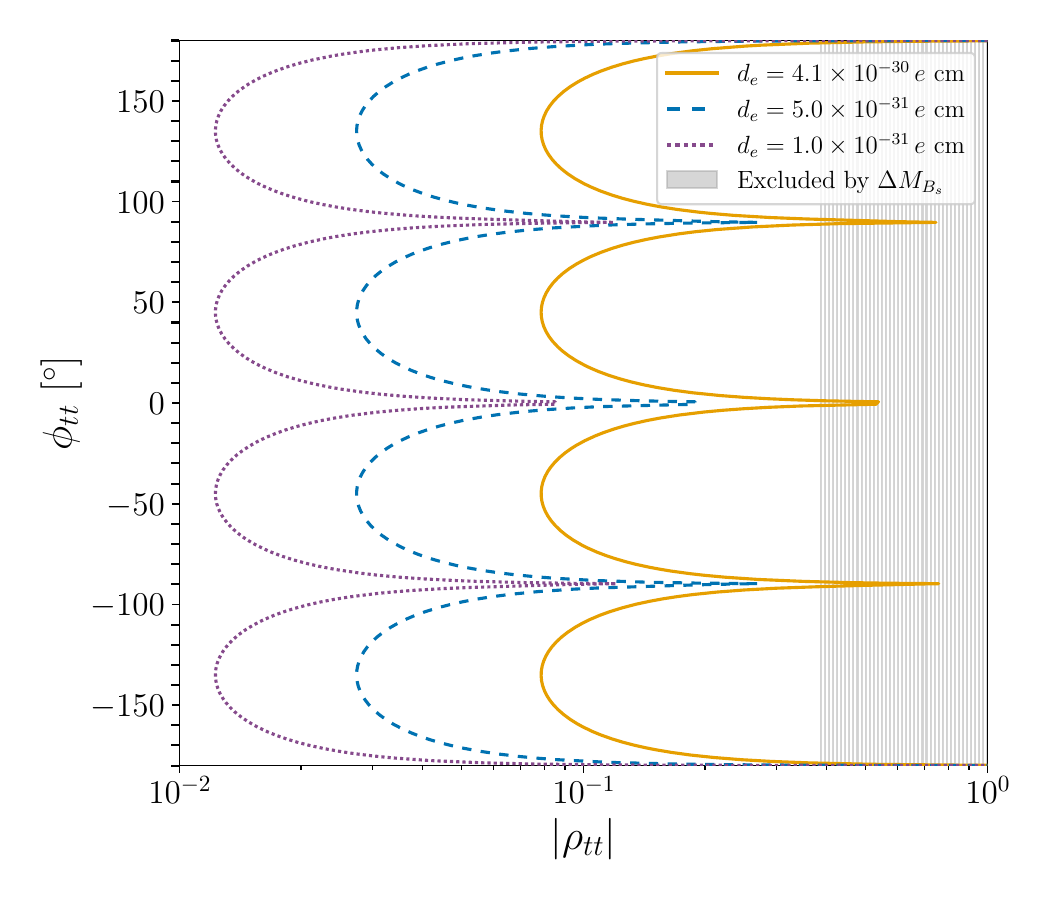}
\caption{The contours of $|d_e|$ are shown, where $|d_e|=4.1\times 10^{-30}~e~\text{cm}$ (solid, orange), $|d_e|=5.0\times 10^{-31}~e~\text{cm}$ (dashed, dark blue), and $|d_e|=1.0\times 10^{-31}~e~\text{cm}$ (dotted, purple). 
The shaded region in gray is excluded by the measurement of $\Delta M_{B_s}$.
We take $m_A=m_{H^\pm}=382$ GeV, $\Lambda_2=1.0$, and $m_H$ is chosen to have $m_{h_1}=125.0$ GeV. }
\label{fig:de_rtt_ptt_r0}
\end{figure}

While the analytic expression is useful for illustrating the parameter dependence of $d_e$, its numerical accuracy is limited. 
We therefore evaluate the electron EDM without employing such simplifications.
In Fig.~\ref{fig:de_rtt_ptt_r0}, the contours of $|d_e|$ are shown, where $|d_e|=4.1\times 10^{-30}~e~\text{cm}$ (solid, orange), $|d_e|=5.0\times 10^{-31}~e~\text{cm}$ (dashed, dark blue), and $|d_e|=1.0\times 10^{-31}~e~\text{cm}$ (dotted, purple). 
The shaded region in gray is excluded by the measurement of $\Delta M_{B_s}$, where our theoretical calculation is based on Refs.~\cite{Diaz:2005rv,Iguro:2017ysu}.
We take $m_A=m_{H^\pm}=382$ GeV, $\Lambda_2=1.0$, and $m_H$ is numerically adjusted to realize $m_{h_1}=125.0$ GeV. 
As discussed above, the electron EDM vanishes not only for $\phi_{tt}=0^\circ$ and $\pm180^\circ$ but also for $\pm90^\circ$.
Regardless of $\phi_{tt}$, however, $|\rho_{tt}|\gtrsim 0.3$ is ruled out by the experimental data of $\Delta M_{B_s}$.
Other than the region where $\sin2\phi_{tt}=0$, the current electron EDM experiment excludes $|\rho_{tt}|\gtrsim 0.1$.
If future experiments achieve a sensitivity to the electron EDM at the level of $d_e = 5.0 \times 10^{-31}~e~\mathrm{cm}$, values of $|\rho_{tt}| \simeq 0.03$ will become testable. The prospects can be further extended towards $|\rho_{tt}| \simeq 0.01$ if the sensitivity reaches $d_e = 1.0 \times 10^{-31}~e~\mathrm{cm}$.

\begin{figure}[t]
\centering
\includegraphics[width=7.5cm]{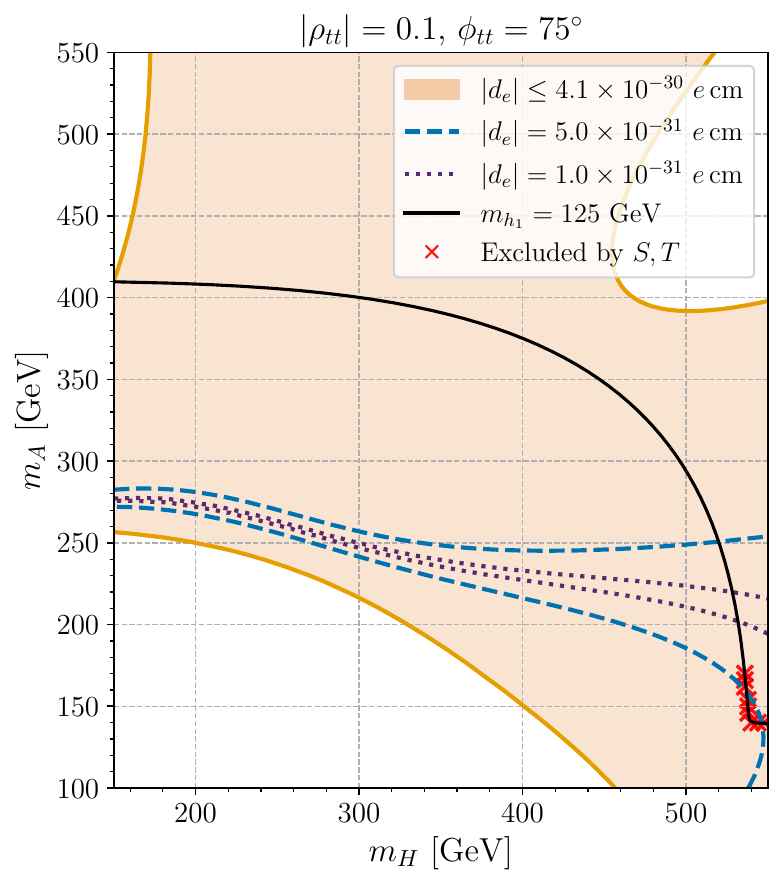}\hspace{0.5cm}
\includegraphics[width=7.5cm]{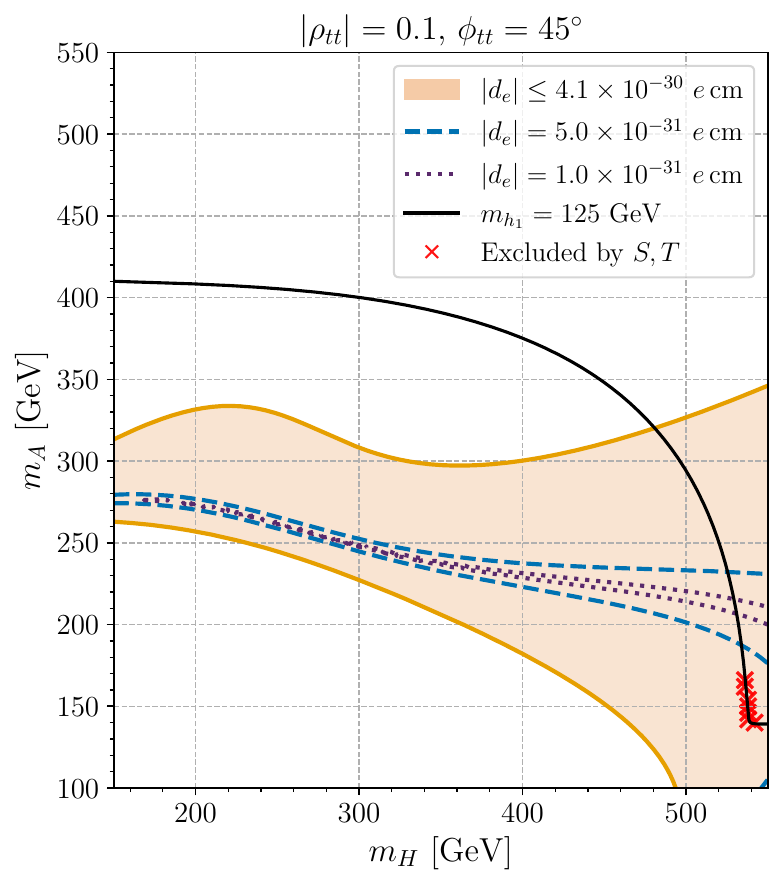}
\caption{
Electron EDM $|d_e|$ in the $(m_H,m_A)$ plane for $\phi_{tt}=75^\circ$ (left) and $45^\circ$ (right).
The orange-shaded region indicates $|d_e| \le 4.1\times10^{-30}~e\,\mathrm{cm}$.
The dark-blue dashed and purple dotted curves show the contours $|d_e|=5.0\times10^{-31}~e\,\mathrm{cm}$ and $|d_e|=1.0\times10^{-31}~e\,\mathrm{cm}$, respectively.
The solid black curve denotes $m_{h_1}=125~\mathrm{GeV}$; cross marks along this curve indicate points excluded by the electroweak $S$–$T$ fit at 95\% CL.
We fix $\Lambda_2=1.0$.
}
\label{fig:de_mH_mA_rtt0p1}
\end{figure}

In Fig.~\ref{fig:de_mH_mA_rtt0p1}, the dependence of $m_H$ and $m_A$ on $|d_e|$ is shown for $\phi_{tt} = 75^\circ$ (left) and $\phi_{tt} = 45^\circ$ (right). The shaded orange region corresponds to the JILA bound $|d_e| < 4.1 \times 10^{-30}~e~\mathrm{cm}$, 
while the contours of $|d_e| = 5.0 \times 10^{-31}~e~\mathrm{cm}$ and $|d_e| = 1.0 \times 10^{-31}~e~\mathrm{cm}$ are indicated by the dashed dark blue and dotted purple curves, respectively. 
The solid black line denotes the $m_{h_1} = 125.0~\mathrm{GeV}$ contour, and the red crosses along this line indicate points excluded by the $S$ and $T$ parameter constraints from the electroweak fit at 95\%~CL~\cite{ParticleDataGroup:2024cfk}. 
In this analysis, we employ the $S$, $T$, and $U$ formulas given in Ref.~\cite{Haber:2010bw}.

It should be noted that constraints from $B$-physics processes, such as $B \to X_s \gamma$, are not sufficient to exclude the parameter region with $|\rho_{tt}| = 0.1$. 

Among the LHC constraints, the signal strength of $h_1 \to \gamma\gamma$ ($\mu_{\gamma\gamma}$) and direct searches for the charged Higgs boson are potentially relevant in the presence of $\rho_{tt}$. Due to the charged Higgs loop contribution, $\mu_{\gamma\gamma}$ can be as small as 0.9, which lies within the $2\sigma$ range of the ATLAS measurement, 
$\mu_{\gamma\gamma}^\mathrm{ATLAS} = 1.04^{+0.10}_{-0.09}$~\cite{ATLAS:2022tnm}, but outside the $2\sigma$ range of the CMS result, $\mu_{\gamma\gamma}^\mathrm{CMS} = 1.12^{+0.09}_{-0.09}$~\cite{CMS:2021kom}. Additional data will therefore be required to reach a definitive conclusion. If the charged Higgs mass is below about 160 GeV, $\mu_{\gamma\gamma}$ can fall outside the two-sigma ATLAS range; however, this region is already excluded by the electroweak precision fit.  

ATLAS places a 95\% CL upper bound on $\sigma(pp \to \bar{t}bH^+)\,\mathrm{Br}(H^+ \to t\bar{b}) < (3.6\,\text{--}\,0.036)\,\mathrm{pb}$ for $m_{H^\pm} \in [200,2000]$~GeV~\cite{ATLAS:2021upq}. Similarly, CMS reports a 95\% CL upper bound on $\sigma(pp \to tbH^\pm)\,\mathrm{Br}(H^\pm \to tb) < (9.25\,\text{--}\,0.005)\,\mathrm{pb}$ for $m_{H^\pm} \in [200,3000]$~GeV~\cite{CMS:2020imj}. The parameter space with $|\rho_{tt}| = 0.1$ is consistent with these bounds.  

Other LHC constraints depend sensitively on additional Yukawa couplings, such as $\rho_{bb}$, $\rho_{\tau\tau}$, etc., which are set to zero in the present analysis; hence no strong bounds arise from them.

We find that the Higgs mass dependence on the electron EDM is nontrivial. This may be attributed to the complicated Higgs mass function of the off-diagonal elements of $\mathcal{M}_N^2$ with the mass relation $(m_H^4+m_A^4+2m_{H^\pm}^4)^{1/4}\simeq 540$ GeV.
For $\phi_{tt}=75^\circ$, points away from the $m_H\simeq m_A=382$ GeV remain consistent with the current electron EDM bound. For $\phi_{tt}=45^\circ$, on the other hand, even though the nearly-degenerated mass point $m_H\simeq m_A=382$ GeV is ruled out by the current electron EDM data, the allowed region could be found when $m_H$ increases and, correspondingly, $m_A$ decreases.
We also find that future electron EDM experiments can further probe the Higgs mass spectrum, squeezing the allowed region to $m_H\gtrsim 500$ GeV and $m_A\lesssim 250$ GeV for $\sin2\phi_{tt}\neq 0$. 

\begin{figure}[t]
\center
\includegraphics[width=8.5cm]{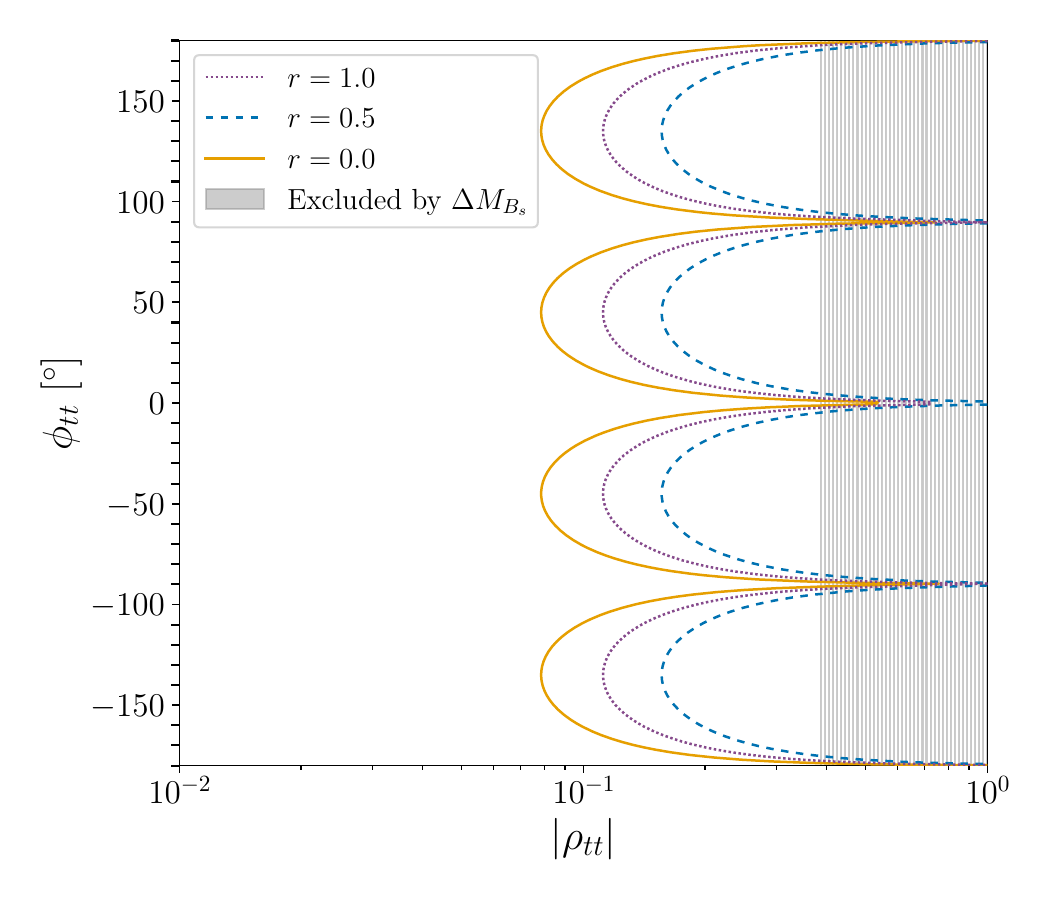}
\includegraphics[width=7.8cm]{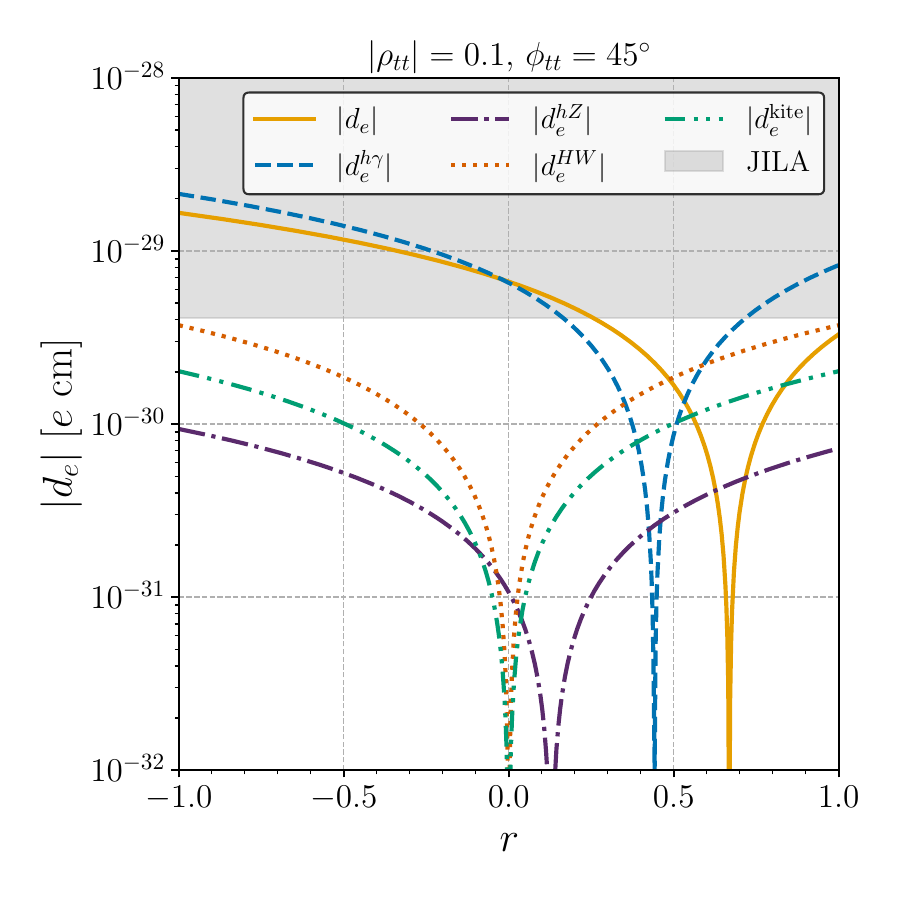}
\caption{
In the left panel, the contours represent $|d_e| = 4.1 \times 10^{-30}~e~\mathrm{cm}$ for $r = 0.0$ (solid, orange), $r = 0.5$ (dashed, dark blue), and $r = 1.0$ (dotted, purple), with $\mathrm{Re}\,\rho_{ee} = -r y_e \mathrm{Re}\,\rho_{tt}/y_t$ and $\mathrm{Im}\,\rho_{ee} = r y_e \mathrm{Im}\,\rho_{tt}/y_t$. 
The gray shaded region is excluded by the measurement of $\Delta M_{B_s}$.
The same input parameters as in Fig.~\ref{fig:de_rtt_ptt_r0} are used.
In the right panel, the decomposition of $|d_e|$ (solid, orange) into $|d_e^{h\gamma}|$ (dashed, dark blue), $|d_e^{hZ}|$ (dash–dot, purple), $|d_e^{HW}|$ (dotted, red), and $|d_e^\mathrm{kite}|$ (dash–dot–dot, green) is shown for $|\rho_{tt}| = 0.1$ and $\phi_{tt} = 45^\circ$. The shaded region is excluded by the JILA experiment.
}
\label{fig:de_rtt_ptt_r}
\end{figure}

%
%
%
\subsection{Case (ii) $\rho_{tt}\neq0$ and $\rho_{ee}\neq0$}
In this case, all the contributions come in. 
After straightforward calculation, the factor $\mathcal{X}_t^{h\gamma}$ in the presence of the nonzero $\rho_{ee}$ is found to be
\begin{align}
\mathcal{X}_t^{h\gamma} & = 
\frac{1}{2}\mathrm{Re}\rho_{tt}\mathrm{Im}\rho_{ee} \Big[-y_t\hat{a}\Delta f_{12}+f(\tau_{th_1})+g(\tau_{th_3}) \Big]
+\frac{1}{2}\mathrm{Re}\rho_{ee}\mathrm{Im}\rho_{tt} \Big[-y_e\hat{b}\Delta f_{23}+f(\tau_{th_3})+g(\tau_{th_1}) \Big]\nonumber\\
& \quad
+\frac{\hat{c}}{2}\Big[\mathrm{Re}\rho_{tt}\mathrm{Re}\rho_{ee}+\mathrm{Im}\rho_{tt}\mathrm{Im}\rho_{ee} \Big]
\mathrm{Re}\rho_{tt}\mathrm{Im}\rho_{tt}(-\Delta f_{13}+\Delta g_{13}) \nonumber\\
&\quad
+\frac{y_e}{2}\mathrm{Re}\rho_{tt}\mathrm{Im}\rho_{tt}\Big[-\hat{a}\Delta g_{12}+\hat{b}\Delta g_{23} \Big].
\end{align}
where $\Delta f_{ij}=f(\tau_{th_i})-f(\tau_{th_j})$.
Now, $\mathcal{X}_t^{h\gamma}$ contains terms proportional to $\mathrm{Re}\rho_{tt}\,\mathrm{Im}\rho_{ee}$ and $\mathrm{Re}\rho_{ee}\mathrm{Im}\rho_{tt}$, in addition to $\mathrm{Re}\rho_{tt}\mathrm{Im}\rho_{tt}$, 
that is, $\mathrm{Im}\rho_{tt}^2$, $\mathrm{Im}(\rho_{tt}\rho_{ee})$, and $\mathrm{Im}(\rho_{tt}\rho_{ee}^*)$. 
If $\rho_{ee}$ is proportional to either $\rho_{tt}$ or $\rho_{tt}^*$, the electron EDM becomes simply proportional to $\mathrm{Im}\,\rho_{tt}^2(=|\rho_{tt}|^2\sin2\phi_{tt})$. 
We adopt the parametrization of $\rho_{ee}$ used in Ref.~\cite{Fuyuto:2019svr}, which takes the form
\begin{align}
\rho_{ee} = -r \frac{y_e}{y_t}\,\rho_{tt}^*,
\label{rhoee_para}
\end{align}
where $r$ is a real parameter, which measures how close the flavor structure between the two $\rho_{ff}$ matrices is to the SM Yukawa structure, namely,  $|\rho_{ee}/\rho_{tt}|=r|y_e/y_t|$.
In this parametrization, all terms proportional to $\mathrm{Im}(\rho_{tt}\rho_{ee})$ vanish.

As is the $\mathcal{X}_t^{h\gamma}$ case, $\mathcal{X}_W^{h\gamma}$ and $\mathcal{X}_{H^\pm}^{h\gamma}$ are estimated as
\begin{align}
\mathcal{X}_W^{h\gamma}
& \simeq -\frac{1}{\sqrt{2}}
\Big[
	\hat{a}\mathrm{Re}\rho_{tt}\mathrm{Im}\rho_{ee}\Delta\mathcal{J}_{W12}^\gamma
	+\hat{b}\mathrm{Re}\rho_{ee}\mathrm{Im}\rho_{tt}\Delta\mathcal{J}_{W23}^\gamma
\Big], \\
\mathcal{X}_{H^\pm}^{h\gamma} 
& \simeq \frac{1}{\sqrt{2}}
\Big[
	\mathrm{Im}\rho_{ee}
	\big\{
		\mathrm{Re}\Lambda_7 \mathcal{I}(\tau_{H^\pm h_1})
		-\hat{a}\Lambda_3\mathrm{Re}\rho_{tt}\Delta\mathcal{I}_{12} 
		+\hat{c}\mathrm{Im}\Lambda_7\mathrm{Re}\rho_{tt}\mathrm{Im}\rho_{tt}\Delta\mathcal{I}_{13} 
	\big\} \nonumber\\
&\hspace{1.5cm}
	-\mathrm{Re}\rho_{ee}
	\big\{
		\mathrm{Im}\Lambda_7 \mathcal{I}(\tau_{H^\pm h_3})
		+\hat{b}\Lambda_3\mathrm{Im}\rho_{tt}\Delta\mathcal{I}_{23} 
		+\hat{c}\mathrm{Re}\Lambda_7\mathrm{Re}\rho_{tt}\mathrm{Im}\rho_{tt}\Delta\mathcal{I}_{13} 
	\big\}
\Big],
\end{align}
where $\Delta \mathcal{J}_{Wij}^\gamma=\mathcal{J}_W^\gamma(m_{h_i})-\mathcal{J}_W^\gamma(m_{h_j})$ and $\Delta \mathcal{I}_{ij}=\mathcal{I}(\tau_{H^\pm h_i})-\mathcal{I}(\tau_{H^\pm h_j})$.
With the parametrization of Eq.~(\ref{rhoee_para}), together with the expressions for $\Lambda_7$ given in Eqs.~(\ref{ReL7}) and (\ref{ImL7}), $\mathcal{X}_W^{h\gamma}$ and $\mathcal{X}_{H^\pm}^{h\gamma}$ are also proportional to $|\rho_{tt}|^2 \sin 2\phi_{tt}$. 

We may infer the relative magnitude and sign between $(d_e^{h\gamma})_t$ and  $(d_e^{h\gamma})_W$ from the analysis of $h_i \to \gamma\gamma$, since the Barr-Zee diagrams shown in Fig.~\ref{fig:BZ} reduce to the Feynman diagrams for $h_i \to \gamma\gamma$ once the electron line is removed. 
In the latter process, the $W$-loop contribution is dominant, followed by the top-loop contribution, and the two interfere destructively. 
An identical or closely similar pattern is likewise observed in the electron EDM.
In addition, $(d_e^{h\gamma})_{H^\pm}$ can also be relevant, as $\Lambda_3$ may be of $\mathcal{O}(1)$ in magnitude, as follows from the tree-level relation $\Lambda_3 = 2 m_{H^\pm}^2 / v^2$.

For $d_e^{hZ}$ and $d_e^{HW}$ defined in Eq.~(\ref{de_decomposed}), their CP-violating dependences are analogous to that of $d_e^{h\gamma}$, precisely as required by the rephasing invariance structure. 

Now we turn to the kite contributions. As shown in Ref.~\cite{Altmannshofer:2020shb}, the kite contribution can be numerically comparable to the Barr-Zee diagrams. 
To ensure consistency with the $W$-loop Barr-Zee diagrams in Eq.~(\ref{de_decomposed}), 
which are based on Ref.~\cite{Abe:2013qla}, the kite contributions that should be included are~\cite{Altmannshofer:2020shb}
\begin{align}
\frac{d_e^\mathrm{kite}}{e} = 
\frac{\sqrt{2} \alpha_\mathrm{em} G_F m_e}{64\pi^3}
\Big[ \delta_\mathrm{kite}^\mathrm{NC} + \delta_\mathrm{kite}^\mathrm{CC} \Big],
\end{align}
where the explicit expressions for $\delta_\mathrm{kite}^\mathrm{NC}$ and 
$\delta_\mathrm{kite}^\mathrm{CC}$ are provided in Appendix~\ref{app:edms}. 
As shown in Eqs.~(\ref{de_kite_NC}) and (\ref{de_kite_CC}), these contributions are 
proportional to $g_{h_i\bar{e}e}^P g_{h_iVV}$, as is $(d_e^{h\gamma})_W$, 
implying that $d_e^\mathrm{kite} \propto |\rho_{tt}|^2 \sin 2\phi_{tt}$. 
Thus, the electron EDM can be written as
\begin{align}
d_e = d_e^\mathrm{BZ} + d_e^\mathrm{kite} 
    = |\rho_{tt}|^2 \sin 2\phi_{tt} (x + r y),
\end{align} 
where $x$ and $y$ are functions of the heavy Higgs boson masses.

The vanishing of the electron EDM with an $\mathcal{O}(1)$ CP-violating phase is realized if $r = -x/y$. 
Following Ref.~\cite{Fuyuto:2019svr}, we refer to this as \textit{structured cancellation} when $r = \mathcal{O}(1)$, since the hierarchy among the $\rho_{ff}$ elements echoes the SM Yukawa structure, $|\rho_{ee}/\rho_{tt}| \simeq y_e/y_t$; otherwise, it is referred to as \textit{unstructured cancellation}. 
The electron EDM cancellation in the context of $\rho_{tt}$-EWBG~\cite{Fuyuto:2019svr} corresponds to 
structured cancellation, whereas that in the context of $\rho_{bb}$-EWBG~\cite{Modak:2020uyq} corresponds 
to unstructured cancellation. 
In the following, we numerically evaluate $d_e$ to identify which type of cancellation is realized.

In the left panel of Fig.~\ref{fig:de_rtt_ptt_r}, the contours of $|d_e| = 4.1 \times 10^{-30}~e~\text{cm}$ are shown in the 
$(|\rho_{tt}|, \phi_{tt})$ plane for $r = 0.0$ (solid, orange), $r = 0.5$ (dashed, dark blue), and $r = 1.0$ (dotted, purple). 
We set $m_A = m_{H^\pm} = 382~\text{GeV}$ and $\Lambda_2 = 1.0$, while $m_H$ is adjusted such that $m_{h_1} = 125.0~\text{GeV}$. 
The gray shaded region is excluded by the $\Delta M_{B_s}$ measurement, as indicated in 
Fig.~\ref{fig:de_rtt_ptt_r0}. The electron EDM vanishes at $|\phi_{tt}| = 90^\circ$, in addition to 
$|\phi_{tt}| = 0^\circ$ and $180^\circ$, as expected. Moreover, for $r \neq 0$, partial cancellation occurs in $d_e$, arising mainly from 
the destructive interference between $d_e^{h\gamma}$ and other contributions that grow with $r$. 

In the right panel, the decomposition of $|d_e|$ (solid, orange) into $|d_e^{h\gamma}|$ (dashed, dark blue), $|d_e^{hZ}|$ (dash–dot, purple), $|d_e^{HW}|$ (dotted, red), and $|(d_e)_\mathrm{kite}|$ (dash–dot–dot, green) is shown as functions of $r$ for $|\rho_{tt}| = 0.1$ and $\phi_{tt} = 45^\circ$. The shaded region is excluded by the JILA experiment. As shown, $d_e^{HW}$ and $(d_e)_\mathrm{kite}$ increase with $r$ and can become comparable to $d_e^{h\gamma}$. At the cancellation point $r \simeq 0.67$, the cancellation primarily occurs among $(d_e^{h\gamma})_t(<0)$, $(d_e^{h\gamma})_W(>0)$, $(d_e^{h\gamma})_{H^\pm}(>0)$, $(d_e^{HW})_{H^\pm/h_i}(<0)$, and $(d_e^\mathrm{CC})_\mathrm{kite}(<0)$. This figure also demonstrates that the cancellation is possible only for positive $r$ in the parametrization of Eq.~(\ref{rhoee_para}).

\begin{figure}[H]
\centering
\includegraphics[width=7.7cm]{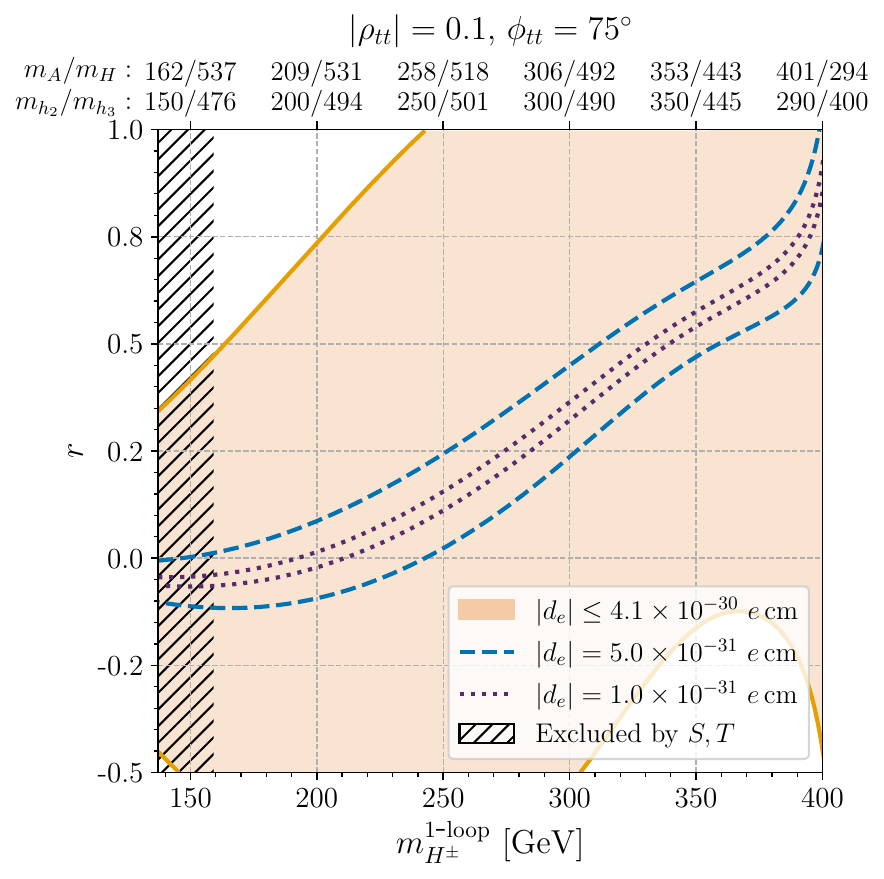} \hspace{0.5cm}
\includegraphics[width=7.7cm]{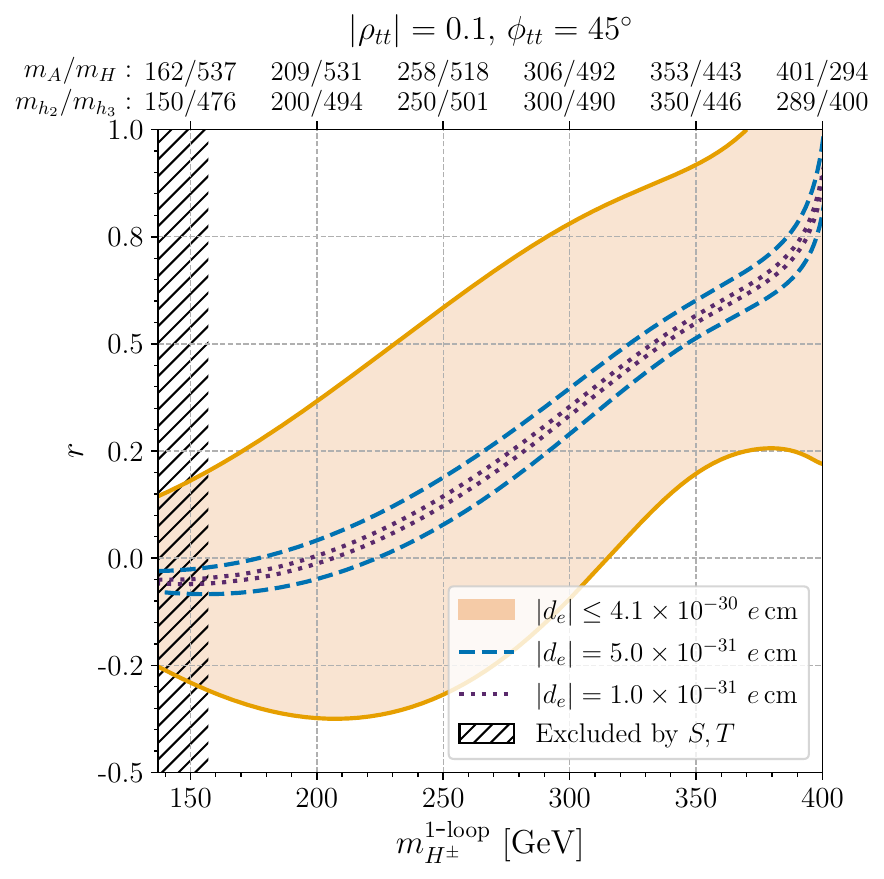} \\[0.5cm]
\includegraphics[width=7.7cm]{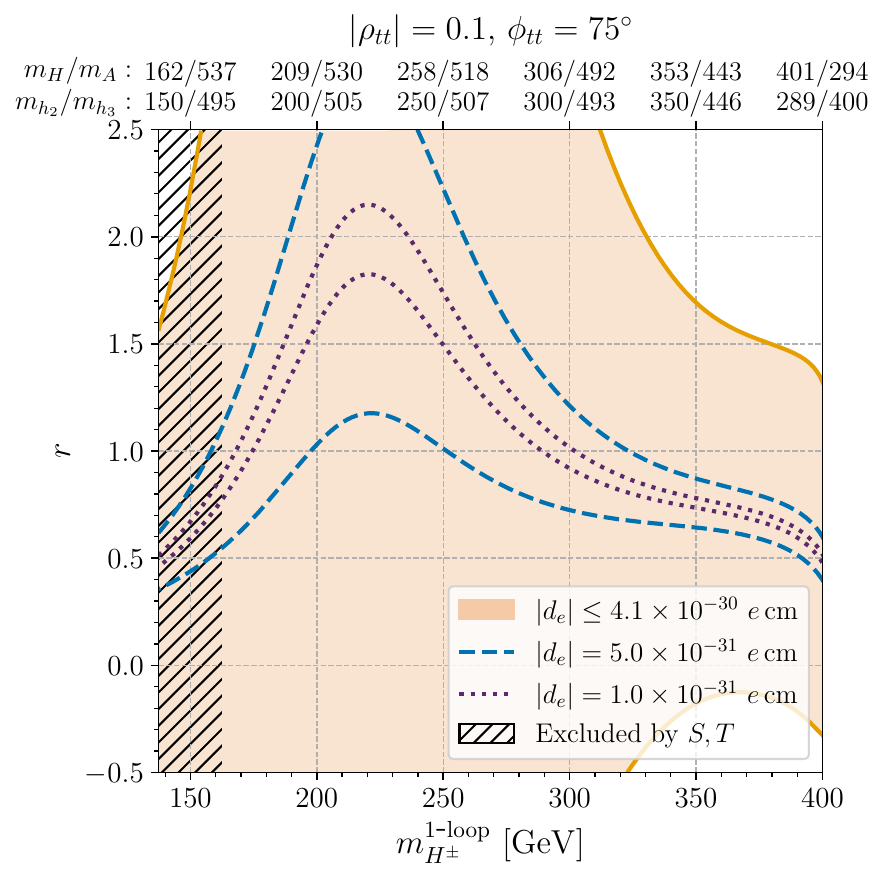} \hspace{0.5cm}
\includegraphics[width=7.7cm]{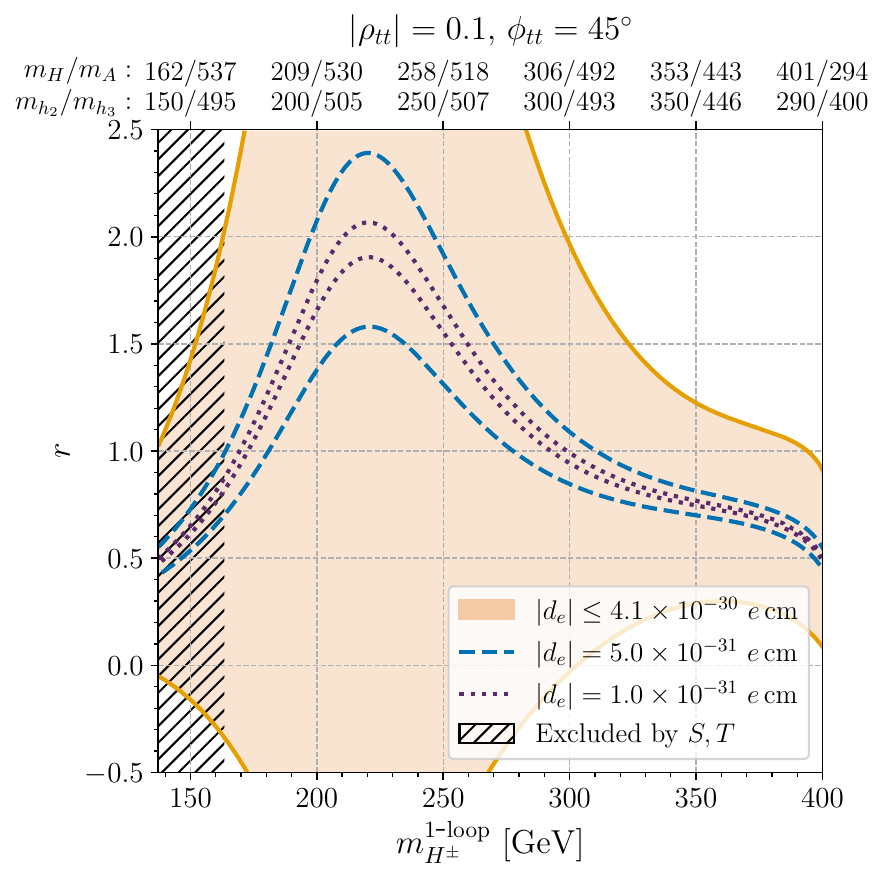}
\caption{Electron EDM $|d_e|$ on the $(m_{H^\pm}^{\text{1\mbox{-}loop}},r)$ plane for $|\rho_{tt}|=0.1$ with $\phi_{tt}=75^\circ$ (left) and $45^\circ$ (right), where $r$ is a real parameter defined via $\rho_{ee}=-r\rho_{tt}^*(y_e/y_t)$. Upper row: custodial spectrum defined at tree level, $m_A=m_{H^\pm}$; lower row: twisted custodial spectrum defined at tree level, $m_H=m_{H^\pm}$. The common $x$-axis uses the one-loop–corrected charged-Higgs mass $m_{H^\pm}^{\text{1\mbox{-}loop}}$ (Eq.~(\ref{mch_1L})). The orange shading indicates $|d_e|\le 4.1\times 10^{-30}\,e\,\mathrm{cm}$; dark-blue dashed and purple dotted contours mark $|d_e|=5.0\times 10^{-31}$ and $1.0\times 10^{-31}\,e\,\mathrm{cm}$, respectively. Hatched regions are excluded by the electroweak $S$–$T$ fit at 95\%~CL. Secondary $x$-axis ticks annotate $m_A, m_H$ (upper) and $m_{h_2}, m_{h_3}$ (lower).}
\label{fig:de_Mch_r_mAmHeqmch}
\end{figure}

We now present the dependence of the Higgs mass spectrum on the electron EDM by considering both the custodially symmetric ($m_A=m_{H^\pm}$) and twisted custodially symmetric ($m_H=m_{H^\pm}$) cases.
Fig.~\ref{fig:de_Mch_r_mAmHeqmch} shows the $|d_e|$ contours on the $(m_{H^\pm}^\mathrm{1\mbox{-}loop}, r)$ plane for $|\rho_{tt}|=0.1$ with $\phi_{tt}=75^\circ$ (left) and $45^\circ$ (right). 
The horizontal axis uses the one-loop–corrected charged-Higgs mass $m_{H^\pm}^{\mathrm{1\mbox{-}loop}}$ as given in Eq.~(\ref{mch_1L}), while the scan itself is performed by varying the \emph{tree-level} heavy-Higgs masses ($m_H, m_A, m_{H^\pm}$) subject to the above custodial relations; $m_{H^\pm}^{\mathrm{1\mbox{-}loop}}$ is thus a \emph{derived} quantity used only as a physically meaningful plotting variable, not an independent input. 
Its accessible range is restricted by the requirement $m_{h_1}=125~\text{GeV}$. 
On the secondary top $x$-axis we display $(m_A,m_H)$ for the upper row and $(m_{h_2},m_{h_3})$ for the lower row. 
In the custodially symmetric case, we have $m_{H^\pm}^\mathrm{1\mbox{-}loop}\simeq m_{h_2}\sim m_A$ for $m_{H^\pm}^\mathrm{1\mbox{-}loop}<382~\text{GeV}$ and $m_{H^\pm}^\mathrm{1\mbox{-}loop}\simeq m_{h_3}\sim m_A$ for $m_{H^\pm}^\mathrm{1\mbox{-}loop}>382~\text{GeV}$; in the twisted custodially symmetric case, $m_{H^\pm}^\mathrm{1\mbox{-}loop}\simeq m_{h_2}\sim m_H$ for $m_{H^\pm}^\mathrm{1\mbox{-}loop}<382~\text{GeV}$ and $m_{H^\pm}^\mathrm{1\mbox{-}loop}\simeq m_{h_3}\sim m_H$ for $m_{H^\pm}^\mathrm{1\mbox{-}loop}>382~\text{GeV}$. 
The colors and line styles of the $|d_e|$ contours match those in Fig.~\ref{fig:de_mH_mA_rtt0p1}, and the hatched regions are excluded by the electroweak $S$–$T$ fit at $95\%$~CL.
In the custodially symmetric case, the electron EDM vanishes in the low-$m_{H^\pm}$ region for $r\simeq0$ (as already seen in Fig.~\ref{fig:de_mH_mA_rtt0p1}); for $r>0$, this cancellation region shifts to larger values of $m_{H^\pm}$. For $m_{H^\pm}^\mathrm{1\mbox{-}loop}\simeq 400$ GeV, we find $|d_e|=0$ at $r\simeq 0.9$. 
In contrast, in the twisted custodially symmetric case, no cancellation is observed at $r\le0$, whereas solutions with $|d_e|=0$ appear for $r>0$. In all cases, such cancellations are possible only when $r=\mathcal{O}(1)$, namely, structured cancellation, analogous to the situation discussed in Ref.~\cite{Fuyuto:2019svr}.

Before closing, some comments are in order. 
\begin{itemize}
\item Other EDMs, such as those of the neutron and mercury, are also subject to experimental upper bounds~\cite{Abel:2020pzs,Graner:2016ses}:
\begin{align}
|d_n| &< 1.8\times 10^{-26}~e~\mathrm{cm}\quad (90\%~\mathrm{CL}), \\
|d_\mathrm{Hg}| &< 7.4\times 10^{-30}~e~\mathrm{cm}\quad (95\%~\mathrm{CL}).
\end{align}
Furthermore, by combining the upper limit on the mercury EDM with nuclear calculations, Ref.~\cite{Sahoo:2016zvr} inferred an indirect bound on the proton EDM, $|d_p| < 2.1\times 10^{-25}~e~\mathrm{cm}$.
We find that these EDMs remain below the current experimental sensitivities in the parameter space under consideration, and thus impose no additional constraints.

\item It is known that the SI-2HDM has a relatively low cutoff scale at which a Landau pole emerges. This scale is typically identified as the energy scale at which one of the Higgs quartic couplings diverges. For example, in the softly $Z_2$-broken SI-2HDM~\cite{Fuyuto:2015jha}, the Landau pole scale is found to be 6.3~TeV. In the g2HDM, however, the Higgs quartic couplings depend on the choice of basis, such as the Higgs/Georgi basis or the generic basis. For instance, $\Lambda_1$ is given by $\sum_i c_i \lambda_i$, where $\lambda_i$ are the Higgs quartic couplings and $c_i$ are model-dependent coefficients in the generic basis. Therefore, the scale at which $|\Lambda_1| \to \infty$ generally differs from the scale at which $|\lambda_i| \to \infty$ for some $\lambda_i$. We are not aware of a basis-independent study of the Landau pole scale in the literature. In the Higgs/Georgi basis adopted throughout our analysis, it is found that the Landau pole scale in the explored parameter space lies in the range 3--10~TeV using one-loop $\beta$-functions~\cite{Ferreira:2009jb}
\end{itemize}

\section{Conclusion}\label{sec:conclusion}
We have analyzed the Higgs mass spectrum and couplings in the SI-g2HDM with explicit CP violation, using the GW method. At tree level, the scalar potential exhibits a flat direction, resulting in a massless scalar and the absence of CP-even and CP-odd mixings in the neutral Higgs mass matrix. At the one-loop level, radiative electroweak symmetry breaking generates both a nonzero mass and CP-violating mixings for the massless scalar, yielding $m_{h_1} = 125$~GeV.

The GW flatness conditions fix the complex Higgs quartic coupling $\Lambda_7$ in terms of the extra Yukawa couplings $\rho_{ff}$ and the heavy Higgs masses. Assuming only the extra top Yukawa coupling $\rho_{tt}$ is significant, we numerically evaluated the Higgs couplings to gauge bosons and fermions. For a typical parameter set, the lightest neutral Higgs boson exhibits SM-like couplings to gauge bosons, $\left|g_{h_1VV}\right| \approx 1$, while the heavier states satisfy $\left|g_{h_2VV}\right|, \left|g_{h_3VV}\right| \lesssim \mathcal{O}(10^{-2})$. In contrast, the charged Higgs couplings, $\left|g_{h_1H^+W^-}\right| = \mathcal{O}(10^{-2})$ and $\left|g_{h_{2,3}H^+W^-}\right| \approx 1$, indicate stronger interactions between the charged Higgs bosons, the heavier neutral Higgs bosons, and the $W$ boson. The top Yukawa couplings satisfy $\left|g^S_{h_1\bar{t}t}\right| \approx 0.7$, $\left|g^P_{h_1\bar{t}t}\right| < 10^{-3}$, and $\left|\kappa_t\right| \approx 1$, with $\left|\tilde{\kappa}_t\right| = \mathcal{O}(10^{-3})$, consistent with current LHC constraints.

We also investigated CP-violating contributions to the electron EDM arising from $\rho_{tt}$ and an additional complex electron Yukawa coupling $\rho_{ee}$. In the absence of $\rho_{ee}$, the EDM is proportional solely to $\mathrm{Im}\,\rho_{tt}^2$, a direct consequence of the GW flatness conditions [Eqs.~(\ref{ReL7}), (\ref{ImL7})], which distinguish the SI-g2HDM from the ordinary g2HDM. 
These conditions also imply a nontrivial dependence of the neutral Higgs mass spectrum on the EDM. In the custodially symmetric case, parameter regions consistent with $m_{h_1} = 125$~GeV and suppressed EDMs arise when $m_A (\simeq m_{h_2}) < m_H (\simeq m_{h_3})$; for example, vanishing EDM points appear for $m_H > 500$~GeV and $200~\mathrm{GeV} < m_A < 250~\mathrm{GeV}$ 
(see Figs.~\ref{fig:de_mH_mA_rtt0p1} and \ref{fig:de_Mch_r_mAmHeqmch}). 
By contrast, no such cancellation region exists in the twisted custodially symmetric case. 
When a nonzero electron Yukawa coupling is further introduced through the parametrization $\rho_{ee} = -r \rho_{tt}^*(y_e/y_t)$,
which preserves the proportionality of the EDM to $\mathrm{Im}\,\rho_{tt}^2$, a nontrivial cancellation region emerges in both custodially and twisted custodially symmetric mass spectra for $r = \mathcal{O}(1)$ (see Fig.~\ref{fig:de_Mch_r_mAmHeqmch}).

Since CP violation induced by $\rho_{tt}$ has been proposed as a viable source of the baryon asymmetry~\cite{Fuyuto:2017ewj}, the possibility of realizing successful EWBG within the SI-g2HDM is both important and of broad interest. A detailed study of this aspect is left for future work. The fact that the electron EDM can be suppressed in this model---similarly to the ordinary g2HDM~\cite{Fuyuto:2019svr}---further strengthens the plausibility of the scenario. At the same time, the predictive mass relations enforced by scale invariance restrict the parameter space more strongly than in the ordinary g2HDM, thereby making future experimental tests particularly decisive.

To investigate the cancellation regions in greater detail, complementary constraints from other EDM measurements and collider searches will be essential. A systematic study including additional Yukawa couplings, such as $\rho_{bb}$ and $\rho_{\tau\tau}$, also represents an important direction for future work.

\begin{acknowledgments}
This work has been supported by the Charles University Research Centre program No. UNCE/24/SCI/016.
E.S. thanks Tanmoy Modak and the participants of the workshop ‘Mori-no-Miyako Workshop on Particle Physics – Yasuhiro Okada Festa’ for valuable comments and discussions.
\end{acknowledgments}
\appendix
\section{Higgs couplings to fermions}\label{app:Hcoup}
After performing the bi-unitary transformations of the fermion fields given in 
Eq.~(\ref{biUnitary}), the Yukawa interactions in Eq.~(\ref{Lag_Y}) take the following form 
in the mass eigenstate basis.
\begin{align}
-\mathcal{L}_{\rm Yukawa}
&= \sum_{f=u,d,e}m_i^{(f)}\bar{f}_if_i\nonumber\\
&= \bar{u}_i
\left[
	\frac{y_i^u}{\sqrt{2}}O_{1a}\delta_{ij}
	+\frac{1}{\sqrt{2}}\rho_{ij}^{u}(O_{2a}-iO_{3a})P_R
	+\frac{1}{\sqrt{2}}\rho_{ij}^{u\dagger}(O_{2a}+iO_{3a})P_L
\right]u_j h_a \nonumber\\
&\quad+ \bar{d}_i
\left[
	\frac{y_i^d}{\sqrt{2}}O_{1a}\delta_{ij}
	+\frac{1}{\sqrt{2}}\rho_{ij}^{d}(O_{2a}+iO_{3a})P_R
	+\frac{1}{\sqrt{2}}\rho_{ij}^{d\dagger}(O_{2a}-iO_{3a})P_L	
\right]d_j h_a \nonumber\\
&\quad+ \bar{e}_i
\left[
	\frac{y_i^e}{\sqrt{2}}O_{1a}\delta_{ij}
	+\frac{1}{\sqrt{2}}\rho_{ij}^{e}(O_{2a}+iO_{3a})P_R
	+\frac{1}{\sqrt{2}}\rho_{ij}^{e\dagger}(O_{2a}-iO_{3a})P_L	
\right]e_j h_a \nonumber\\
&\quad
+\bigg[
\bar{u}_i\
\Big\{
	(V_{\rm CKM}\rho^{d})_{ij}P_R-(\rho^{u\dagger}V_{\rm CKM})_{ij}P_L
\Big\}d_jH^+
+\bar{\nu}_i\rho^{e}_{ij}P_Re_jH^+
+\mathrm{H.c.}
\bigg],
\label{Lag_Y_mass}
\end{align}
where $O$ is the orthogonal matrix defined in Eq.~(\ref{defO}), $V_\mathrm{CKM}$ denotes Cabibbo-Kobayashi-Maskawa matrix~\cite{Cabibbo:1963yz,Kobayashi:1973fv}, 
and the operators $P_{L,R} = (1 \mp \gamma_{5})/2$ are the standard chirality projectors constructed from the Dirac gamma matrices.

Let us parametrize the Higgs couplings to the fermions in Eq.~(\ref{Lag_Y_mass}) as
\begin{align}
-\mathcal{L}_{h_i\bar{f}_jf_k} 
& = h_i\bar{f}_j\left(g_{h_i\bar{f}_jf_k}^RP_R+g_{h_i\bar{f}_jf_k}^LP_L\right)f_k
= h_i\bar{f}_j\left(g_{h_i\bar{f}_jf_k}^S+i\gamma_5g_{h_i\bar{f}_jf_k}^P\right)f_k,  
\label{ffphi} \\
-\mathcal{L}_{H^+\bar{f}_{\uparrow} f_{\downarrow}} 
&=  H^+\bar{f}_\uparrow \left(g_{H^+\bar{f}_{\uparrow} f_{\downarrow}}^RP_R+g_{H^+\bar{f}_{\uparrow} f_{\downarrow}}^LP_L\right)f_\downarrow +\mathrm{H.c.} \nonumber  \\
 &= H^+\bar{f}_\uparrow \left(g_{H^+\bar{f}_{\uparrow} f_{\downarrow}}^S+i\gamma_5g_{H^+\bar{f}_{\uparrow} f_{\downarrow}}^P\right)f_\downarrow +\mathrm{H.c.}, \label{ffHpm}
\end{align}
where $f_\uparrow=u,\nu$ and $f_\downarrow=d,e$, and 
\begin{align}
g_{\phi\bar{f}_Af_B}^S = \frac{g_{\phi\bar{f}_Af_B}^R+g_{\phi\bar{f}_Af_B}^L}{2},\quad
g_{\phi\bar{f}_Af_B}^P = -i\frac{g_{\phi\bar{f}_Af_B}^R-g_{\phi\bar{f}_Af_B}^L}{2},
\end{align}
with $\phi=h_i$, $H^+$.
For the diagonal couplings of the neutral Higgs bosons to fermions, the interactions can be written in a simpler form shown in Eqs.~(\ref{gS}) and (\ref{gP}).

\section{One-loop corrections to the Higgs masses}\label{app:1LSigma}
The one-loop corrections to the diagonal elements of $\mathcal{M}_N^2$ can be obtained by taking $p^2=0$ in the following expressions:
\begin{align}
\bar{\Sigma}_{h_1'}(p)
&= \frac{-1}{16\pi^2}
\bigg[
-4\sum_fN_C^f
\bigg\{
	\frac{1}{2}\big( g_{h\bar{f}_if_j}^Sg_{h\bar{f}_jf_i}^S+g_{h\bar{f}_if_j}^Pg_{ h\bar{f}_jf_i}^P \big)\Big( \bar{A}(m_{f_i})+\bar{A}(m_{f_j})   \nonumber \\
&\hspace{7.5cm}
	+(m_{f_i}^2+m_{f_j}^2-p^2)\bar{B}_0(p^2; m_{f_i}, m_{f_j}) \Big) \nonumber\\
&\hspace{3.5cm}
	+m_{f_i}m_{f_j}\big( g_{h\bar{f}_if_j}^Sg_{h\bar{f}_jf_i}^S-g_{h\bar{f}_if_j}^Pg_{h\bar{f}_jf_i}^P \big) \bar{B}_0(p^2; m_{f_i}, m_{f_j})
\bigg\} \nonumber \\
&\hspace{2cm}
	+\frac{1}{2}\sum_{\phi=H,A,H^\pm}
	n_\phi
	\Big(
	\lambda_{hh\phi\phi}\bar{A}(m_\phi) 
	+\lambda_{h\phi\phi}^2\bar{B}_0(p^2; m_\phi, m_\phi)
	\Big)
\bigg],\\
\bar{\Sigma}_{h_2'}(p) 
&=\frac{-1}{16\pi^2}\bigg[
-4\sum_fN_C^f
\bigg\{
	\frac{1}{2}\big( g_{H\bar{f}_if_j}^Sg_{H\bar{f}_jf_i}^S+g_{H\bar{f}_if_j}^Pg_{H\bar{f}_jf_i}^P \big)\Big( \bar{A}(m_{f_i})+\bar{A}(m_{f_j})   \nonumber \\
&\hspace{7.5cm}
	+(m_{f_i}^2+m_{f_j}^2-p^2)\bar{B}_0(p^2; m_{f_i}, m_{f_j}) \Big) \nonumber\\
&\hspace{3.5cm}
	+m_{f_i}m_{f_j}\big( g_{H\bar{f}_if_j}^Sg_{H\bar{f}_jf_i}^S-g_{H\bar{f}_if_j}^Pg_{H\bar{f}_jf_i}^P \big) \bar{B}_0(p^2; m_{f_i}, m_{f_j})
\bigg\} \nonumber \\
&\hspace{1.5cm}
	+\frac{1}{2}\lambda_{hhHH}\bar{A}(m_h)+\frac{1}{2}\lambda_{HHHH}\bar{A}(m_H)
	+\frac{1}{2}\lambda_{HHAA}\bar{A}(m_A)+\lambda_{HHH^+H^-}\bar{A}(m_{H^\pm}) \nonumber\\
&\hspace{1.5cm}
	+\frac{1}{2}\lambda_{HHH}^2\bar{B}_0(p^2; m_H, m_H)
	+\frac{1}{2}\lambda_{HAA}^2\bar{B}_0(p^2; m_A, m_A) 
	+\lambda_{HH^+H^-}^2\bar{B}_0(p^2; m_{H^\pm}, m_{H^\pm}) \nonumber\\
&\hspace{1.5cm}	
	+\lambda_{hHH}^2\bar{B}_0(p^2; m_h, m_H)
	+\lambda_{HG^0A}^2\bar{B}_0(p^2; m_{G^0}, m_A) \nonumber \\
&\hspace{1.5cm}
	+\lambda_{HHA}^2\bar{B}_0(p^2; m_H, m_A)
	+2\lambda_{HH^\pm G^\mp}^2\bar{B}_0(p^2; m_{G^\pm}, m_{H^\pm})
	\bigg], \label{Sigma_H_IDM}\\
\bar{\Sigma}_A(p)
&=\frac{-1}{16\pi^2}
\bigg[
-4\sum_fN_C^f
\bigg\{
	\frac{1}{2}\big( g_{A\bar{f}_if_j}^Sg_{A\bar{f}_jf_i}^S+g_{A\bar{f}_if_j}^Pg_{A\bar{f}_jf_i}^P \big)\Big( \bar{A}(m_{f_i})+\bar{A}(m_{f_j})   \nonumber \\
&\hspace{7.5cm}
	+(m_{f_i}^2+m_{f_j}^2-p^2)\bar{B}_0(p^2; m_{f_i}, m_{f_j}) \Big) \nonumber\\
&\hspace{3.5cm}
	+m_{f_i}m_{f_j}\big( (g_{A\bar{f}_if_j}^Sg_{A\bar{f}_jf_i}^S-g_{A\bar{f}_if_j}^Pg_{A\bar{f}_jf_i}^P\big) \bar{B}_0(p^2; m_{f_i}, m_{f_j})
\bigg\} \nonumber \\
&\hspace{1.5cm}
	+\frac{1}{2}\lambda_{hhAA}\bar{A}(m_h)+\frac{1}{2}\lambda_{HHAA}\bar{A}(m_H)
	+\frac{1}{2}\lambda_{AAAA}\bar{A}(m_A)+\lambda_{AAH^+H^-}\bar{A}(m_{H^\pm}) \nonumber\\
&\hspace{1.5cm}	
	+\lambda_{hAA}^2\bar{B}_0(p^2; m_h, m_A)
	+\lambda_{HAA}^2\bar{B}_0(p^2; m_H, m_A)	
	+\lambda_{HG^0A}^2\bar{B}_0(p^2; m_{G^0}, m_H) \nonumber\\
&\hspace{7cm}
	+2\lambda_{AH^+G^-}\lambda_{AH^-G^+}\bar{B}_0(p^2; m_{G^\pm}, m_{H^\pm}) \nonumber \\
&\hspace{1.5cm}
	+\frac{1}{2}\lambda_{AAA}^2\bar{B}_0(p^2; m_A, m_A)+\frac{1}{2}\lambda_{HHA}^2\bar{B}_0(p^2; m_H, m_H)
	+\lambda_{AH^+H^-}^2\bar{B}_0(p^2; m_{H^\pm}, m_{H^\pm}) 
\bigg], \\
\bar{\Sigma}_{H^\pm}(p)
&= \frac{-1}{16\pi^2}
\bigg[
-4\sum_fN_C^f
\bigg\{
	\frac{1}{2}\big( |g_{H^+\bar{f}_\uparrow f_\downarrow }^S|^2+|g_{H^+\bar{f}_\uparrow f_\downarrow }^P|^2 \big)\Big( \bar{A}(m_{f_\uparrow })+\bar{A}(m_{f_\downarrow})   \nonumber \\
&\hspace{7.5cm}
	+(m_{f_\uparrow}^2+m_{f_\downarrow}^2-p^2)\bar{B}_0(p^2; m_{f_\uparrow}, m_{f_\downarrow}) \Big) \nonumber\\
&\hspace{3.5cm}
	+m_{f_\uparrow}m_{f_\downarrow}\big( |g_{H^+\bar{f}_\uparrow f_\downarrow }^S|^2-|g_{H^+\bar{f}_\uparrow f_\downarrow }^P|^2 \big) \bar{B}_0(p^2; m_{f_\uparrow}, m_{f_\downarrow})
\bigg\} \nonumber \\
&\hspace{1.5cm}
	+\frac{1}{2}\lambda_{hhH^+H^-}\bar{A}(m_h)+\frac{1}{2}\lambda_{HHH^+H^-}\bar{A}(m_H)
	+\frac{1}{2}\lambda_{AAH^+H^-}\bar{A}(m_A) \nonumber\\
&\hspace{1.5cm}	
	+\lambda_{H^+H^-H^+H^-}\bar{A}(m_{H^\pm})
	+\lambda_{hH^+H^-}^2\bar{B}_0(p^2; m_h, m_{H^\pm})
	+\lambda_{HH^+H^-}^2\bar{B}_0(p^2; m_H, m_{H^\pm})
	 \nonumber\\
&\hspace{1.5cm}
	+\lambda_{HH^\pm G^\mp}^2\bar{B}_0(p^2; m_{G^\pm}, m_{H})
	+\lambda_{AH^+G^-}\lambda_{AH^-G^+}\bar{B}_0(p^2; m_{G^\pm}, m_{A}) \nonumber \\
&\hspace{1.5cm}
	+\lambda_{AH^+H^-}^2\bar{B}_0(p^2; m_A, m_{H^\pm})
\bigg],
\end{align}
where the self-energies are regularized in the $\overline{\mathrm{MS}}$-scheme. 
The self-couplings are defined as
\begin{align}
\lambda_{\phi_i\phi_j\phi_k} = \frac{\partial^3 V_0}{\partial \phi_i \partial \phi_j \partial \phi_k},\quad
\lambda_{\phi_i\phi_j\phi_k\phi_\ell} = \frac{\partial^4 V_0}{\partial \phi_i \partial \phi_j \partial \phi_k \partial \phi_\ell}.
\end{align}
The couplings $g_{\phi\bar{f}_if_j}^{S,P}$ and $g_{H^+\bar{f}_\uparrow f_\downarrow }^{S,P}$ are defined in Eqs.~(\ref{ffphi}) and (\ref{ffHpm}).
We take the approximation defined in Eq.~(\ref{diagonalization_tree}) with $s_\gamma=1$ in the one-loop self-energies, as the dropped terms are of two-loop order.
The loop functions $\bar{A}(m)$ and $\bar{B}_0$ are, respectively, given by
\begin{align}
\bar{A}(m) & = -m^2\left(\ln\frac{m^2}{\bar{\mu}^2}-1\right),\\
\bar{B}_0(p^2; m_1,m_2) &= 
-\int_0^1dx~\ln\left[\frac{-x(1-x)p^2+(1-x)m_1^2+xm_2^2-i\epsilon}{\bar{\mu}^2}\right],
\end{align}
where the $i\epsilon~(\epsilon>0$) term is needed when $\bar{B}_0$ has an imaginary part. 
For $p^2=0$, one gets 
\begin{align}
\bar{B}_0(0; m_1, m_2)
&= \frac{-1}{m_1^2-m_2^2}
\left[
	m_1^2\left(\ln\frac{m_1^2}{\bar{\mu}^2}-1\right)-m_2^2\left(\ln\frac{m_2^2}{\bar{\mu}^2}-1\right)
\right].
\end{align}
Note that $\bar{B}_0(0; m,m) = -\ln(m^2/\bar{\mu}^2)$.
The general analytic expressions of $\bar{B}_0$ can be found in Ref.~\cite{Hagiwara:1994pw}.

\section{Electron EDM}\label{app:edms}
We collect formulas for calculating the electron EDM. The relevant interactions are parameterized as
\begin{align}
\mathcal{L}_{A/Z\bar{f}f} &= -eQ_fA_\mu\bar{f}\gamma^\mu f 
-g_ZZ_\mu\bar{f}\gamma^\mu(v_{Z\bar{f}f}-a_{Z\bar{f}f}\gamma_5)f, \\
\mathcal{L}_{W^+\bar{f}f} &= -W^+_\mu\bar{f}_\uparrow\gamma^\mu
	\Big[g_{W^+\bar{f}_\uparrow f_\downarrow}^{L}P_L\Big]f_\downarrow +\mathrm{H.c}, \\
\mathcal{L}_{h_iH^+H^-}&=-\bar{\lambda}_{h_iH^+H^-}vh_iH^+H^-, \\
\mathcal{L}_{ZH^+H^-}&=ig_{ZH^+H^-}^{}\Big\{(\partial_\mu H^-)H^+-H^-(\partial_\mu H^+)\Big\}Z^\mu,
\end{align}
where $e$ is the positron charge, $Q_f$ denotes the electric charge of the fermion $f$, $g_Z=g_2/c_W$, and
\begin{align}
v_{Z\bar{f}f} &= \frac{1}{2}T_f^3-Q_fs^2_W,\quad
a_{Z\bar{f}f} = \frac{1}{2}T_f^3, \\
g_{W^+\bar{q}_\uparrow q_\downarrow}^L 
&= \frac{g_2}{\sqrt{2}}(V_{\rm CKM})_{\uparrow\downarrow},\quad
g_{W^+\bar{\ell}_\uparrow \ell_\downarrow}^L 
= \frac{g_2}{\sqrt{2}}, \\
\bar{\lambda}_{h_iH^+H^-} &= \Lambda_3O_{1i}+\mathrm{Re}\Lambda_7O_{2i}-\mathrm{Im}\Lambda_7O_{3i}, \\
g_{ZH^+H^-}^{} & = e\cot2\theta_W,
\end{align}
with $T_f^3=\pm 1/2$ being the third component of the weak isospin. 

The expressions for $d_e^{hZ}$ and $d_e^{HW}$ defined in Eq.~(\ref{de_decomposed}) are, respectively, given
 by~\cite{Barr:1990vd,Ellis:2008zy,Cheung:2009fc,West:1993tk,Chang:2005ac,Ellis:2010xm,Cheung:2014oaa,Inoue:2014nva,Bowser-Chao:1997kjp,Abe:2013qla,Jung:2013hka,Hou:2021zqq,Hou:2023kho,Altmannshofer:2025nsl,Davila:2025goc}
\begin{align}
\frac{(d_{e}^{hZ})_t}{e} 
& = -\sum_{i}\frac{\alpha_{\text{em}}v_{Z\bar{e}e}v_{Z\bar{t}t}}{4\pi^3m_fs_W^2c_W^2}
\Big[g_{h_i\bar{e}e}^Sg_{h_i\bar{t}t}^P\tilde{F}_A(\tau_{th_i},\tau_{tZ})+g_{h_i\bar{e}e}^Pg_{h_i\bar{t}t}^S\tilde{F}_H(\tau_{th_i},\tau_{tZ}) \Big], \\
\frac{(d_{e}^{hZ})_W}{e}
&=\sum_{i} \frac{\sqrt{2}\alpha_{\text{em}}G_Fvv_{Z\bar{e}e}}{32\pi^3s_W^2} g_{h_i \bar{e}e}^Pg_{h_i VV}^{}\mathcal{J}^Z_W(m_{h_i}), \\
\frac{(d_{e}^{hZ})_{H^\pm}}{e}
&=\sum_{i}\frac{g_Z^{}v_{Z\bar{e}e}g_{ZH^+H^-}v}{128\pi^4m_{H^\pm}^2}
\bar{\lambda}_{h_i H^+H^-}g_{h_i \bar{e}e}^P\Big[\tilde{F}_H(\tau_{H^\pm h_i}, \tau_{H^\pm Z})-\tilde{F}_A(\tau_{H^\pm h_i}, \tau_{H^\pm Z}) \Big], \\
\frac{(d_e^{HW})_{t/b}}{e}
& = -\frac{N_C^f}{256\pi^4m_{H^\pm}^2}\int_0^1dx\left(\frac{Q_t}{x}+\frac{Q_b}{1-x} \right)
J\left(r_{WH^\pm}, \frac{r_{tH^\pm}}{x}+\frac{r_{bH^\pm}}{1-x}\right) \nonumber\\
&\quad \times 
\mathrm{Im}
\bigg[
\Big\{g_{W^+\bar{\nu}e}^{L*}\big(g^S_{H^+\bar{\nu}e}+ ig^P_{H^+\bar{\nu}e}\big)\Big\} \nonumber\\
&\hspace{1.4cm}\times
\Big\{ G_+^{LR} m_t(1-x)^2+G_-^{LR}m_t(1-x) +G_+^{RL} m_bx^2+G_-^{RL}m_bx\Big\}
\bigg],\\
\frac{(d_e^{HW})_{W/h_i}}{e}
&= -\frac{e^2}{(16\pi^2)^2\sqrt{2}vs_W^2}\sum_i\text{Im}\Big[g^S_{H^+\bar{\nu} e}g_{h_iH^+W^-}^*\Big] 
g_{h_iVV}^{}\mathcal{I}_4(m_{h_i}^2, m_{H^\pm}^2),\\
\frac{(d_e^{HW})_{H^\pm/h_i}}{e}
&= \frac{\sqrt{2}}{(16\pi^2)^2v}\sum_i\text{Im}\Big[g^S_{H^+\bar{\nu} e}g_{h_iH^+W^-}^*\Big] 
	\bar{\lambda}_{h_iH^+H^-}\mathcal{I}_5(m_{h_i}^2, m_{H^\pm}^2),
\end{align}
where $\tau_{AB}=m_A^2/m_B^2$, $G_\pm^{AB} = (g^S_{H^+\bar{t}b})^*(g_{W^+\bar{t}b}^A\pm g_{W^+\bar{t}b}^B)
+i(g^P_{H^+\bar{t}b})^*(g_{W^+\bar{t}b}^A\mp g_{W^+\bar{t}b}^B)$, and the loop functions are, respectively, defined as
\begin{align}
f(\tau) &= \frac{\tau}{2}\int_0^1dx~\frac{1-2x(1-x)}{x(1-x)-\tau}\ln\left(\frac{x(1-x)}{\tau}\right),\\
g(\tau) &= \frac{\tau}{2}\int_0^1dx~\frac{1}{x(1-x)-\tau}\ln\left(\frac{x(1-x)}{\tau}\right), \\
\tilde{F}_A(a, b) &= \frac{1}{a-b}\Big[ag(b)-bg(a)\Big], \quad
\tilde{F}_H(a, b) = \frac{1}{a-b}\Big[af(b)-bf(a)\Big], \\
\mathcal{J}_W^V(m_\phi)
&=\frac{2m_W^2}{m_\phi^2-m_V^2}
\bigg[
	-\frac{1}{4}\left\{\left(6-\frac{m_V^2}{m_W^2}\right)+\left(1-\frac{m_V^2}{2m_W^2}\right)\frac{m_\phi^2}{m_W^2} \right\} \nonumber\\
&\hspace{4cm}\times\big(I_1(m_W,m_\phi)-I_1(m_W,m_V) \big) \nonumber\\
&\hspace{2.5cm}
	+\left\{\left(-4+\frac{m_V^2}{m_W^2}\right)+\frac{1}{4}\left(6-\frac{m_V^2}{m_W^2}\right)
	+\frac{1}{4}\left(1-\frac{m_V^2}{2m_W^2}\right)\frac{m_\phi^2}{m_W^2}\right\} \nonumber\\
&\hspace{4cm}\times\big(I_2(m_W,m_\phi)-I_2(m_W,m_V) \big)
\bigg], \\
\mathcal{I}_{4,5}(m_1^2,m_2^2)&=\frac{m_W^2}{m_{H^\pm}^2-m_W^2}
\Big[I_{4,5}(m_W^2, m_1^2)-I_{4,5}(m_2^2,m_1^2) \Big],
\end{align}
with
\begin{align}
I_1(m_1,m_2) & = -2\frac{m_2^2}{m_1^2}f\left(\frac{m_1^2}{m_2^2}\right), \quad
I_2(m_1,m_2)  = -2\frac{m_2^2}{m_1^2}g\left(\frac{m_1^2}{m_2^2}\right), \\
I_4(m_1^2, m_2^2) 
&=\int_0^1dx~(1-x)^2\left[x-4+\frac{m_{H^\pm}^2-m_2^2}{m_W^2}x\right]\nonumber\\
&\hspace{1cm} \times
	\frac{m_1^2}{m_W^2(1-x)+m_2^2x-m_1^2x(1-x)}
	\ln\left(\frac{m_W^2(1-x)+m_2^2x}{m_1^2x(1-x)}\right), \\
I_5(m_1^2, m_2^2) 
&=2\int_0^1dx~	\frac{m_1^2x(1-x)^2}{m_{H^\pm}^2(1-x)+m_2^2x-m_1^2x(1-x)}
	\ln\left(\frac{m_{H^\pm}^2(1-x)+m_2^2x}{m_1^2x(1-x)}\right),\\
J(a, b) &= \frac{1}{a-b}\left[\frac{a}{a-1}\ln a - \frac{b}{b-1}\ln b\right].
\end{align}
Note that $(d_e^{HW})_{t/b}$ is absent in softly-$Z_2$ broken 2HDMs.

In addition to the above contributions, we also include the kite diagrams derived in 
Ref.~\cite{Altmannshofer:2020shb}. To maintain consistency with 
$(d_e^{h\gamma})_W$, $(d_e^{hZ})_W$, and $(d_e^{HW})_{W/h_i}$, which are based on 
Ref.~\cite{Abe:2013qla}, we set $\xi = 1$ when incorporating the kite contributions 
of Ref.~\cite{Altmannshofer:2020shb}.
\begin{align}
\frac{d_e^\mathrm{kite}}{e} =\frac{(d_e^\mathrm{NC})_\mathrm{kite}}{e}+\frac{(d_e^\mathrm{CC})_\mathrm{kite}}{e} = \frac{\sqrt{2}\alpha_\mathrm{em}G_Fm_e}{64\pi^3}\Big[\delta_\mathrm{kite}^\mathrm{NC} +\delta_\mathrm{kite}^\mathrm{CC}\Big],
\end{align}
where
\begin{align}
\delta_\mathrm{kite}^\mathrm{NC} 
&=\sum_i \frac{g_{h_i\bar{e}e}^P g_{h_iVV}^{}}{24s_W^2c_W^2}\frac{\sqrt{2}}{y_e}
\Big[
	3(16v_{Z\bar{e}e}^2-1)K_1(\tau_{Zh_i})+(16v_{Z\bar{e}e}^2+1)K_2(\tau_{Zh_i})
\Big], \label{de_kite_NC} \\
\delta^\mathrm{CC}_\mathrm{kite}
&= \sum_i \frac{g_{h_i\bar{e}e}^P g_{h_iVV}^{}}{4s_W^2}\frac{\sqrt{2}}{y_e}
\Big[
K(\tau_{Wh_i})+F_{\xi=1}(\tau_{Wh_i})-G_{\xi=1}(\tau_{Wh_i})
\Big],
\label{de_kite_CC}
\end{align}
with
\begin{align}
K_1(\tau)&=
\frac{1}{\tau^3}
\bigg[
z^2 + \frac{\pi^2}{6}(1-4\tau)-2\tau^2\ln \tau+\frac{1}{2}(1-4\tau)\ln^2\tau \nonumber\\
&\hspace{1cm}
+2(1-4\tau+\tau^2)\mathrm{Li}_2\Big(1-\frac{1}{\tau}\Big)+\frac{1}{2}(1-6\tau+8\tau^2)\Phi(\tau)
\bigg],\\
K_2(\tau) &=
\frac{1}{\tau}
\bigg[
2\tau(1-4\tau)+\frac{\pi^2}{3}\big(3\tau^2+4\tau^3\big)-2\tau(1+4\tau)\ln \tau \nonumber\\
&\hspace{1cm}
+2\big(1-3\tau^2-4\tau^3\big)\mathrm{Li}_2\!\Big(1-\frac{1}{\tau}\Big)+(1-2\tau-8\tau^2)\Phi(\tau)
\bigg],\\
K(\tau)
&= \frac{2\pi^2}{9}\tau(3+4\tau)+\frac{2}{3}\,(5-8\tau)-\frac{16}{3}(1+\tau)\ln \tau \nonumber\\
&\quad
+\frac{2(3+2\tau-6\tau^3-8\tau^4)}{3\tau^2}\mathrm{Li}_2\!\Bigl(1-\frac{1}{\tau}\Bigr)
+\frac{(1+2\tau)(3-10\tau+\tau^2)}{3\tau^2}\Phi(\tau), \\
F_{\xi=1}(\tau)
&= (3-2 \tau)\Phi(\tau), \\
G_{\xi=1}(\tau)
&= -2+ 2\left[\frac{(1-\tau)^2}{\tau^2}-1\right]\mathrm{Li}_2\!\Bigl(1-\frac{1}{\tau}\Bigr)
+\left[2+\frac{1-4\tau}{\tau^2}\right]\Phi(\tau), \\
\Phi(\tau) &= \frac{2}{\sqrt{1-4\tau}}
\left[
\frac{\pi^2}{6}-\frac{1}{2} \ln^2 \tau+\ln^2\!\left(\frac{1-\sqrt{1-4\tau}}{2}\right)
-2\mathrm{Li}_2\!\left(\frac{1-\sqrt{1-4\tau}}{2}\right)
\right].
\end{align}

%
\bibliography{refs}
%

\end{document}